\documentclass[%
reprint,
 amsmath,amssymb,
 aps,
]{revtex4-1}

\usepackage{graphicx}% Include figure files
\usepackage{dcolumn}% Align table columns on decimal point
\usepackage{bm}% bold math
\usepackage{physics}
\usepackage[mathscr]{euscript}
\usepackage{subfig}
\usepackage{tikz}
\usepackage[title]{appendix}
\usepackage{float}
\usepackage{hyperref}
\hypersetup{
    colorlinks=true,
    linkcolor=blue,
    citecolor=magenta,      
    urlcolor=cyan,
}
 
\begin{document}

\title{Non-equilibrium nature of non-linear optical response: Application to the bulk photo voltaic effect}

\author{Tamoghna Barik $^1$}

\author{Jay D. Sau $^2$}

\affiliation{Condensed Matter Theory Center $^{1,2}$, Department of Physics $^1$ and Joint Quantum Institute $^2$ , \\
University of Maryland, College Park, MD-20742, USA}

\begin{abstract}
The bulk photovoltaic effect is an example of a non-linear optical response that leads to a DC current that is relevant for photo-voltaic applications. In this work, we theoretically study this effect in the presence of electron-phonon interactions. Using the response function formalism we find that the non-linear optical response, in general, contains three operator correlation functions, one of which is not ordered in time. This latter correlator cannot be computed from equilibrium field theory. Using a semiclassical approach instead, we show that the bulk photovoltaic effect can be attributed to the dipole moment of the generated excitons. We then confirm the validity of the semiclassical result (which agrees with the non-interacting result) for non-linear DC response from a quantum master equation approach. From this formalism we find that, in contrast to usual linear response, the scattering rate has a strong implicit effect on the non-linear DC response. Most interestingly, the semiclassical treatment shows that the non-linear DC response for spatially inhomogeneous excitation profiles is strongly non-local and must involve the aforementioned out-of-time-ordered correlators that cannot be computed by equilibrium field theory.
\end{abstract}
\maketitle
\section{Introduction\label{sec:Introduction}}
The fluctuation-dissipation theorem that relates the perturbative response of observables to equilibrium correlators of observables is one of the corner stones of quantum many-body physics. Recently, a number of non-linear phenomena such as non-linear optical response, photogalvanic effect, pump-probe spectroscopy have been used to characterize solid state materials. For example, certain second harmonic generation coefficients have been used to detect a hidden inversion symmetry breaking nematic phase \cite{Harter2017}. Similar correspondence has been shown between second harmonic generation and inversion symmetry breaking in certain topological materials with hexagonal warping term and a gap for quasiparticle excitations \cite{Li2019,Nagaosa2017}. The non-linear response of the DC current in a Weyl material to circularly polarized light, the so-called circular photogalvanic effect has been shown to be quantized \cite{Moore2017}. In fact, the quantized circular photogalvanic response has been directly related to the topological monopole charge of Weyl materials \cite{Moore2017}. \\
The circular photogalvanic effect is one example of a broader class of bulk photovoltaic effects where the  non-linear response to optical excitation results in a DC current. The bulk or anomalous photovoltaic effect, which was discovered experimentally \cite{nature1946} as a large voltage in optically illuminated semiconductors has been attributed to a large number potential sources such as asymmetric scattering from impurities, phonons or electron-electron interactions \cite{Struman1977,Belinicher1978,Strumanrev1980,Struman'sbook,Fridkin'sbook,Spivak2009}.   The bulk photovoltaic effect (BPVE) for linearly polarized light has been argued to be a result of shift currents \cite{Sipe2000} in bulk materials, where the current is generated by the sequential tunneling of electrons to one direction. Theoretically, using a non-linear response formalism, the BPVE was argued to arise intrinsically even in single crystalline materials that break inversion symmetry \cite{Kraut1981}. It was found that the magnitude of this response is directly related to the Berry curvature over parts of the Brillouin zone where the optical perturbation is resonant \cite{Morimoto2016,Kraut1981}. Aside from fundamental interest as a probe of Berry curvature, the BPVE also has potential applications to solar cell devices provided materials with high efficiency can be found \cite{Cook2017,Tan2016,Rangel2017,Rappe2012,Zheng2016,Nakamura2017,Wang2017,Gong2018,Ogawa2017}.\\
Despite the use of characterizing symmetries and topology of materials, the understanding of non-linear optical responses such as the BPVE is quite limited outside the purview of non-interacting systems. As elaborated in Sec. \ref{sec:secordoptresp}, the response function formalism for non-linear effects such as the BPVE contains terms that are neither time-ordered nor anti-time ordered. Such out of time ordered correlated terms that have also been shown to appear in non-linear response of superconductors \cite{Larkin1968} cannot be computed using standard equilibrium Feynman diagram formalism and in this sense are truly non-equilibrium properties.\\
 The utility of the BPVE for solar cell devices ultimately relies on the ability to annihilate photons and convert them to electrical power. However, in the equilibrium diagram formalism, each power of the vector-potential $A(t)$ is associated with either the absorption or emission of a photon. The creation of a DC current at second order in a vector potential would then have to be interpreted as the sequential absorption and emission of a photon. Within this picture, the generation of electrical energy appears quite paradoxical. An electrical current can only generate electrical energy in the presence of a finite voltage that can arise in the presence of a scattering process, which is ignored in the non-linear response theory for shift currents \cite{Sipe2000,Kraut1981,Morimoto2016}. While scattering processes play a crucial role in certain extrinsic mechanisms for the BPVE \cite{Belinicher1978,Struman1977}, the non-linear response mechanism is typically considered to ignore such scattering processes \cite{Sipe2000,Kraut1981,Morimoto2016}. Thus, it is desirable to extend the non-linear response formalism for the BVPE to include such scattering processes in a way that is completely consistent with energy conservation.\\
 \begin{figure}
	\centering
    \subfloat[]{{\includegraphics[width=5cm]{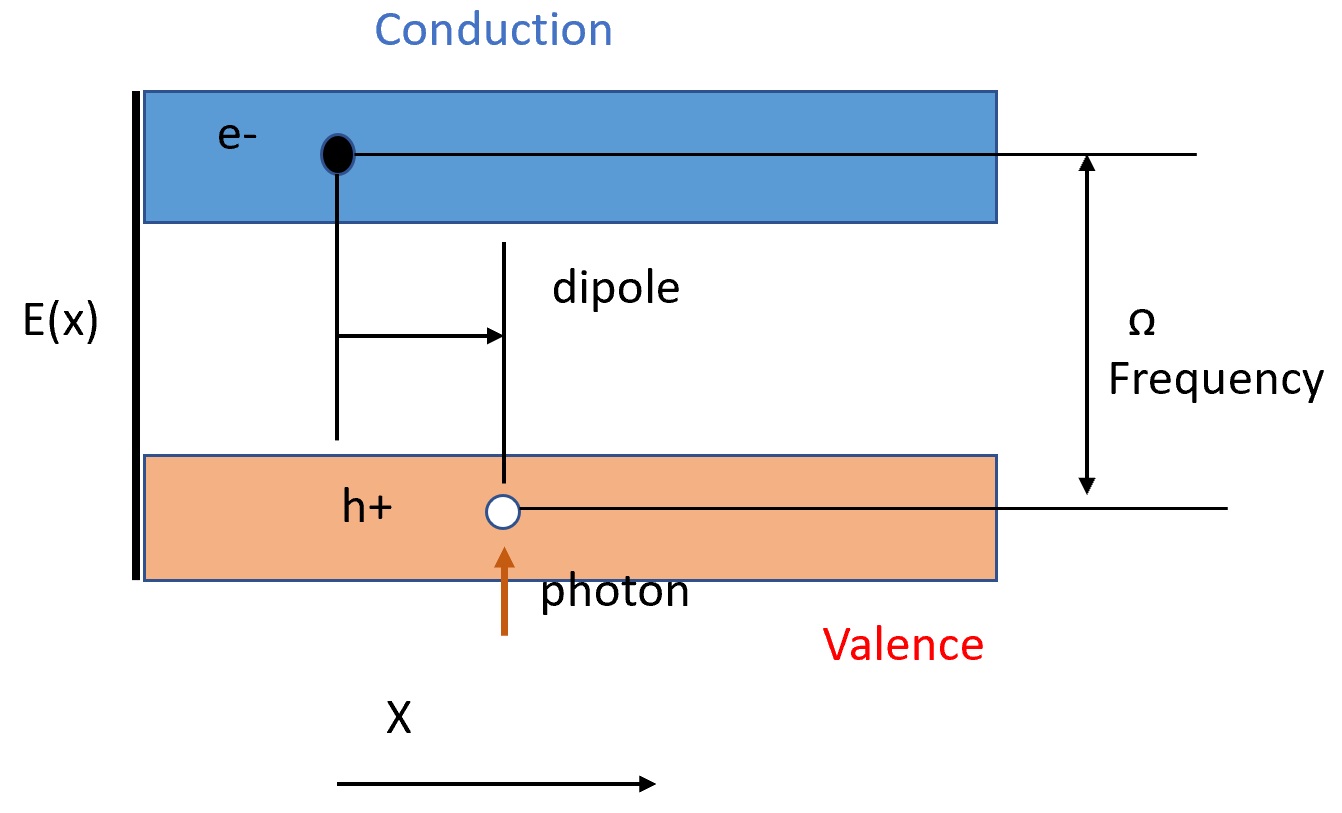}}}%
    \centering
    \subfloat[]{{\includegraphics[width=4cm]{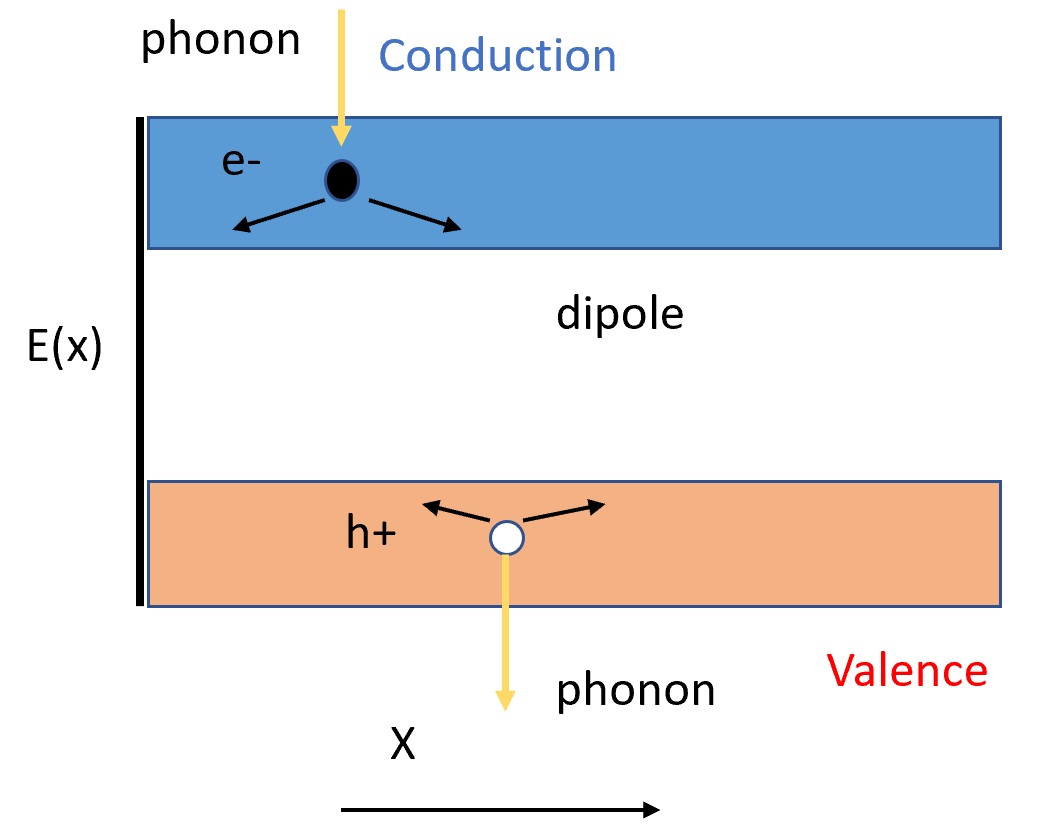}}}
   \caption{Two step process determining DC response: (a)Photon creates e-h pair density with their charge centers separated, generating dipole moment.
(b)Scattering with phonons terminates charge shifting for that e-h pair.}
\label{fig:semiclass}
\end{figure}
 In this work, we systematically study the non-linear response approach to the BVPE in a minimally interacting system, where we include the interaction of the electrons with the phonons in the material. We assume that the electron-phonon interaction is symmetric so as to not directly contribute to the shift current. In Sec. \ref{sec:secordoptresp}, we start by developing a response function formalism for the BPVE and demonstrating the role of out of time-ordered correlators (OTOC) \cite{Larkin1968}. However, the OTOC relevant to this work involves three operators and is not connected to Lyapunov exponents as is the case with the OTOC in Ref. \cite{Larkin1968}. The presence of such correlators invalidates the standard computation using equilibrium Feynman diagrams. In order to gain intuition, we first develop a semi-classical picture of the BPVE (Sec. \ref{sec:semiclass}) based on the processes shown in Fig. \ref{fig:semiclass}. This picture is shown to lead to the conventional expression for the DC current in the case of homogeneous optical excitation. However, the application of inhomogeneous excitation profiles are found to lead to results that cannot be described by equilibrium response functions. Such a semiclassical treatment is only justified in the strong electron-phonon scattering limit relative to the optical excitation. We go beyond this limit in Sec. \ref{sec:quantumapproach} using a Floquet Master equation approach. We find that this approach, in addition to reproducing the known small vector potential limit of the DC response also predicts a linearly scaling DC response in the limit of small electron-phonon scattering. Finally, in Sec. \ref{sec:numerics}, we study the cross-over between the small and large vector potential limit by computing the response numerically within the same formalism for the Rice-Mele model \cite{Rice1982}. 
\section{Second order optical response \label{sec:secordoptresp}}
We consider a generic Hamiltonian for a two band semiconductor coupled to electromagnetic (EM) field, given by
\begin{equation}\label{eq:fullham}
	\hat{H}(t) =  \hat{H}_0 + \int dx A(r) \hat{J}(r) 
\end{equation}
 where $\hat{H}_0$ is the two band system Hamiltonian and $r=(x,t)$ includes both space and time co-ordinates. $\hat{J}(r)$ is the current operator which couples to the EM vector potential, $A(r) = A_0(x) \, e^{\eta t} \left( e^{\iota \Omega t}+e^{-\iota \Omega t} \right) \equiv A_0(x) \, e^{\eta t} \alpha (t)$ where $\eta = 0_+$ is the rate at which the EM field is turned on.\\
 We can expand $\hat{J}(r)$ in powers of $A$ as: $\hat{J}(r) \simeq \hat{J}_0(r) + A(r) \hat{J}_1(r)$. The response $\langle \hat{J}(r) \rangle$, up to second order in $A_0(x)$, can be written as (see Appendix \ref{app:response})
\begin{equation}\label{response}
	\begin{split}
		& \langle \hat{J}(r) \rangle \simeq \langle \hat{J}(r) \rangle _0 + \int _{-\infty} ^{\infty} dx' \, A_0(x') \, \chi _{lin} (r,x') \\
        & + \int _{-\infty} ^{\infty} dx' \, A_0^2(x') \, \chi_{n,2} (r,x') \\
        & + \int _{-\infty} ^{\infty} dx_1 dx_2 \, A_0(x_1)A_0(x_2) \, \chi_{n,3} (r,x_1,x_2) \\
        & +\int _{-\infty} ^{\infty} dx_1 \,  dx_2 \, A_0(x_1) A_0(x_2) \, \chi _{otoc} (r,x_1, x_2) 
	\end{split}
\end{equation}
where $\chi_{lin}$ is linear response given by
\begin{equation}\label{eq:chilin}
	\chi _{lin} (r,x') = \iota \int_{-\infty}^t dt' \, e^{\eta t'}\alpha (t') \left  \langle \comm{\mathscr{J}_0(r')}{\mathscr{J}_0(r)} \right \rangle _0 
\end{equation}
Here the current operators are in interaction picture defined as $\mathscr{J}_{0,1}(r) \equiv e^{\iota \hat{H}_0 t} \hat{J}_{0,1}(r) e^{-\iota \hat{H}_0 t}$. Among the non-linear response functions, $\chi_{n,2}$ is the part of non-linear response that can be viewed as a linear response of the current to a perturbation $A^2_0(r) \, \hat{J}_1(r)$ and is given by:
\begin{equation}\label{eq:chin2}
	\chi_{n,2}(r,x') = \iota \int_{-\infty}^t dt' \, e^{2\eta t'} \alpha ^2(t')\left \langle \comm{\mathscr{J}_1(r')}{\mathscr{J}_0(r)} \right \rangle _0
\end{equation}
The term proportional to $\chi _{n,3}$, while non-linear and involving three current operators, is given by the time-ordered product as \small
\begin{equation}\label{eq:chin3}
	\begin{split}
		& \chi_{n,3}(r,x_1,x_2) \\
        & = -\frac{1}{2}\int_{-\infty}^t dt_1 \, dt_2 e^{\eta (t_1+t_2)} \alpha (t_1) \alpha (t_2) \left  \langle \acomm{\mathcal{T} \{ \mathscr{J}_0(r_1)\mathscr{J}_0(r_2) \}}{\mathscr{J}_0(r)} \right \rangle _0
	\end{split}
\end{equation}
\normalsize $\chi_{otoc}$ involves out of time ordered correlator of cubic current operators given by \small
\begin{equation}\label{eq:chiotoc}
	\begin{split}
		& \chi_{otoc}(r,x_1,x_2) =\int_{-\infty}^t dt_1 \, dt_2 e^{\eta (t_1+t_2)} \alpha (t_1) \alpha (t_2) \left \langle  \tilde{\mathscr{J}}_0(r_1)\tilde{\mathscr{J}}_0(r)\tilde{\mathscr{J}}_0(r_2)  \right \rangle _0
	\end{split}
\end{equation}
\normalsize In these expressions, $\langle .. \rangle_0 $ means expectation value w.r.t. the ground state of the system. Since $\chi_{otoc}(t)$ contains out of time ordered correlator (OTOC), Feynman diagrams cannot be used to evaluate it. One requires Keldysh formalism \cite{Larkin1968} to compute such correlators in presence of any interaction.\\
Another way to see that the zero frequency part of $\langle \hat{J}(r) \rangle $ is likely not computable as a response around equilibrium is to quantize the electromagnetic field in terms of photons (as mentioned in the introduction). Expanding $A(t)=a^\dagger e^{i\omega t}+h.c$ and substituting, A(t) into Eq. \ref{response}, we see that $\langle \hat{J}(r) \rangle _{DC} \propto \langle a a^\dagger \rangle $ , which represents absorption and remission of a photon. This suggests that no energy is absorbed, even though the DC current in the presence of a voltage would do work. This would likely represent a non-equilibrium situation suggesting the need for a non-equilibrium theory. Consequences of this somewhat imprecise argument will become clear from the semiclassical treatment of the DC current in the next section.
    \section{SemiClassical description\label{sec:semiclass}}
In order to understand the absorption of energy more clearly, we include an explicit dissipation process through phonon scattering with a semiclassical approach. In this section we treat electron-phonon scatterings as probabilistic events along with the quantum mechanical evolution of the electronic state due to external EM field. For simplicity we consider two band Hamiltonian for a semiconductor given by
\begin{equation}\label{eq:Bloch}
   		\hat{H}_0 = \sum_k \Delta _k \left( c^{\dagger}_kc_k - v^{\dagger}_k v_k \right)
   \end{equation}
   where $c_k$($v_k$) is conduction (valence) band state operator and $2\Delta _k$ is the band gap.
   \subsection{Shift current in homogeneous excitation\label{subsec:semiclasshomo}}
   The incident EM radiation leads to the creation of electron-hole (e-h) pairs. The time dependent amplitude of such a pair can be extracted from the EM coupling in Eq. \ref{eq:fullham} to be
    \begin{equation}\label{eq:pertshift}
	\begin{split}
    	&  |\delta \Psi_k (t)\rangle _I \simeq  A_0 (J^x_{0,k} - \iota J^y_{0,k}) \frac{\sin \left( \frac{\Omega}{2}-\Delta _k \right)t}{\left( \frac{\Omega}{2}-\Delta _k \right)} |\bar{eh}_k \rangle _I \\
        & \equiv  A_0 |J^{\perp}_{0,k}| e^{\iota \phi_k} \frac{\sin \left( \frac{\Omega}{2}-\Delta _k \right)t}{\left( \frac{\Omega}{2}-\Delta _k \right)} |\bar{eh}_k \rangle _I
	\end{split}
\end{equation}
where $|0\rangle$ is the ground state and $|\bar{eh}_k \rangle \equiv c^{\dagger}_k v_k |0\rangle$ is e-h pair state. $ \vec{J}_0 \cdot \vec{\sigma} \equiv \hat{J}_0 = \partial_A \hat{H(t)}|_{A=0} $ such that  $\vec{\sigma}$ is the vector of Pauli matrices in the band basis (i.e. $\sigma ^z | c_k \rangle =|c_k \rangle$, $\sigma ^z | v_k \rangle =-|v_k \rangle$) and $|J^{\perp}_{0,k}| \equiv \sqrt{(J^x_{0,k})^2 + (J^y_{0,k})^2}$.\\
In non-centrosymmetric materials, the resulting e-h pair can carry a dipole moment. So the electric field generated dipole moment evolves in time accordingly as:
\begin{equation}\label{eq:dipmom}
	\begin{split}
		& P (t) \equiv \sum_k \langle \delta \Psi _k (t) | \left( r_e - r_h \right) | \delta \Psi _k (t) \rangle \\
        & = \frac{A_0^2}{2\pi} \int_{BZ}dk \, \frac{\sin ^2 \left( \frac{\Omega}{2}-\Delta _k \right)t}{\left( \frac{\Omega}{2}-\Delta _k \right)^2} \left( \vec{J}_{0,k} \cross \vec{J}_{1,k} \right) _z
	\end{split}
\end{equation}
where $\vec{J}_{1,k} \equiv -\partial_k \vec{J}_{0,k}$. In this semiclassical approach the coherent evolution under the electric field is interrupted by phonon scatterings modeled as Poisson processes. The coherent evolution time $t$ corresponding to the Poisson process is described by an exponential distribution function as
\begin{equation}\label{eq:poiss}
   	f(t) \equiv \eta \, e^{-\eta t}
\end{equation}
where $\eta$ is the total rate for electron and hole scatterings as detailed in the later part of this subsection. In the weak electric field limit, once the scattering happens the e-h pair disperses equally in either direction and no further dipole moment can be generated. Thus integrating the quantum dipole moment in Eq. \ref{eq:dipmom} with $f(t)$ leads to an average dipole moment generated due to e-h pair, given by
\begin{equation}\label{eq:avgdipmom}
	\begin{split}
		& \langle P \rangle \equiv \int _0 ^{\infty} dt P(t) f(t)= \frac{A_0 ^2}{\pi} \, \int_{BZ}dk \, \frac{\left( \vec{J}_{0,k} \cross \vec{J}_{1,k} \right) _z }{\eta ^2 + \left( \Omega-2\Delta _k \right) ^2} .
	\end{split}
\end{equation}
The DC current can be determined by the average rate of dipole moment generation over the characteristic scattering time, $\tau \equiv \frac{1}{\eta}$. Using this relation we obtain the expression for the shift current to be
\begin{equation}\label{eq:DCsc}
	\begin{split}
		& J_{shift}^{DC} = \frac{\langle P \rangle}{\tau} = \frac{A_0^2 }{ \pi} \int _{BZ} dk \,  \frac{ \eta}{(\Omega - 2\Delta _k)^2 +\eta ^2} (\vec{J}_{0,k} \cross \vec{J}_{1,k})_z .
	\end{split}
\end{equation}
Here $\eta$ is a physical energy scale corresponding to electron phonon scattering strength. Considering the limit of $\lim_{\eta \to 0} \frac{\eta}{(\Omega - 2\Delta _k)^2 +\eta ^2} = \pi \delta (\Omega - 2\Delta _k) $ in Eq. \ref{eq:DCsc} we obtain
\begin{equation}\label{eq:DCsclim}
	\lim _{\eta \to 0} J_{shift}^{DC} = A_0 ^2 \int _{BZ} \, dk \, ( \vec{J}_{0,k} \cross \vec{J}_{1,k} )_z \delta (\Omega - 2 \Delta _k)
\end{equation}
which matches with the known expression for shift current given in \cite{Moore2017} (see Appendix \ref{app:berryconnection}).\\
The semiclassical approach presented above, is valid in the limit of strong e-ph coupling or weak electric field ($A_0|J^{\perp}_{0,k}|\ll\eta$) where the scattering processes are instantaneous events with time scale $\tau \ll \frac{1}{A_0|J^{\perp}_{0,k}|}$. In Sec. \ref{sec:quantumapproach} we will use a more quantum mechanical approach to study the case of longer scattering times $ \tau $. \\
Let us now consider in detail the scattering of electrons by phonons following the coherent exciton generation discussed above. Assuming the phonon scattering matrix elements are momentum independent, the momenta and the corresponding velocities of the electrons and holes are uncorrelated. The resulting motion of the electrons and holes are diffusive starting from the average position. The electrons and holes can recombine and emit a phonon. A current is generated if the electron and hole annihilate with a pair that is different from the way they were created. This is the notion of a shift current. However, if the electron and hole of the generated pair recombine with each other before they can diffuse away, for local electron-hole recombination, the recombined excitons do not have any dipole moment. Therefore, in calculating the shift current, the average dipole moment of the exciton, remains unchanged right until the exciton merges into the electron-hole gas.
\subsection{Diffusion current response for inhomogeneous excitations\label{subsec:inhomexcitation}}
\begin{figure}
	\vspace{.2in}
	\centering
    \includegraphics[width=7cm]{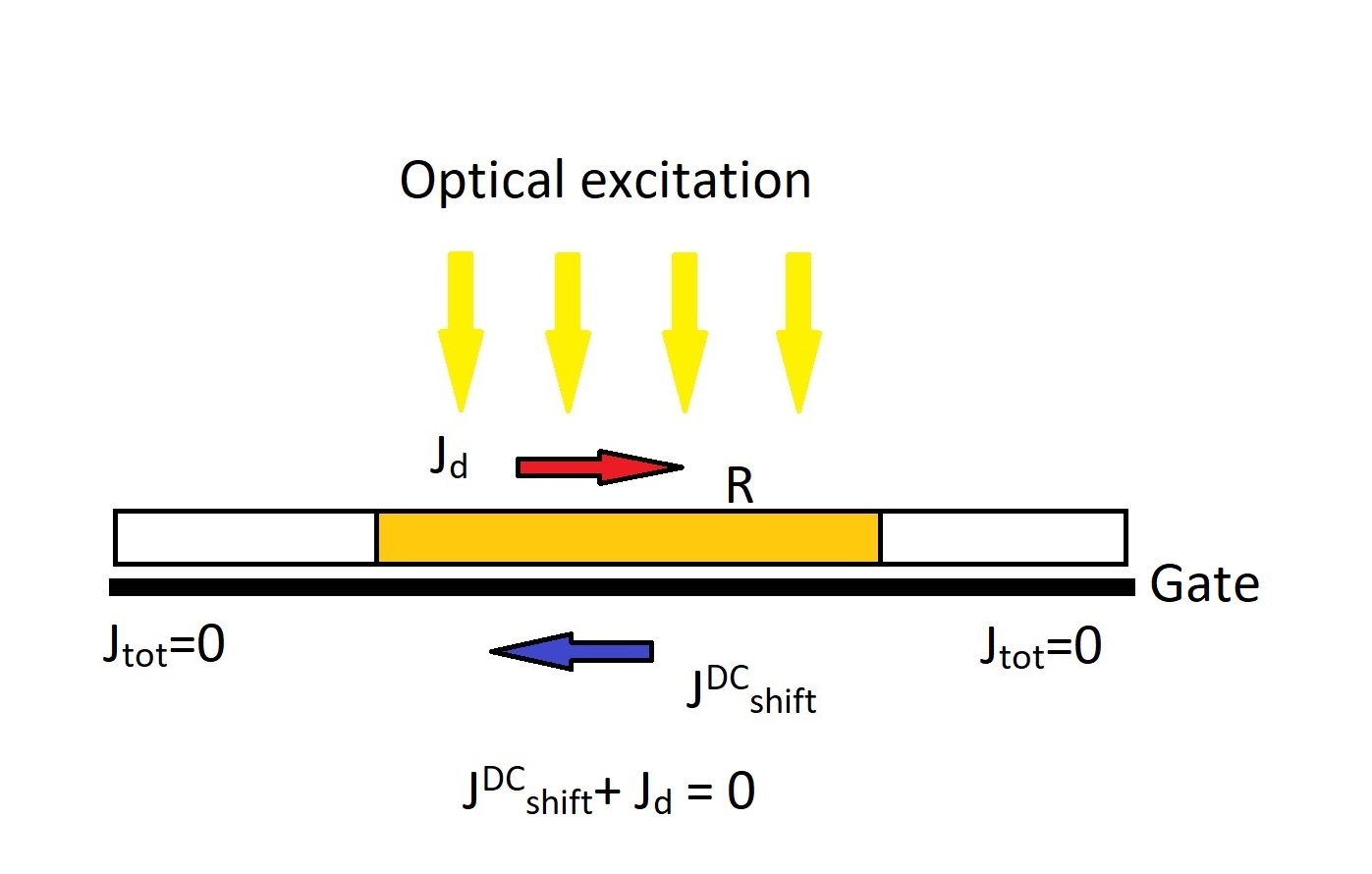}
   \caption{In the optically excited region, $R$, shift current, $J^{DC}_{shift}$, is countered by the diffusion current, $J^d$, so that the total current is divergence-less in the steady state. A gate in proximity can be used to measure the charge density profile capacitively and hence the shift current in steady state.}
   \label{fig:inhom}
\end{figure}
Let us now consider the response of current to an inhomogeneous optical excitation profile applied to the system which is often the case in practical setups \cite{Struman'sbook}. We assume that the intensity of the optical excitation is non-zero over a finite region $R$ in Fig. \ref{fig:inhom} and vanishes outside. We note that at finite temperature, there is finite (although exponentially small as temperature vanishes) carrier density due to thermal excitations on top of EM excited electron and holes. Recombination with these thermally excited carriers leads to a finite lifetime for both electrons and holes, even though the total charge density is conserved. The shift current $J_{shift}^{DC}=\alpha A^2_0(x)$, in the case of an inhomogeneous intensity $A^2_0(x)$, varies in space. This combined with the conservation of charge leads to an inhomogeneous charge density. The gradient in charge leads to a diffusion current $J^d=-D\partial_x\rho$, where the diffusivity of charge $D=\sigma/\chi$ and $\sigma$ and $\chi$ are the conductivity and compressibility of the unperturbed carriers. The density profile is determined from the equation of charge conservation as
\begin{equation}\label{eq:eqnforrho}
	\partial_t \rho (x,t) = -\partial_x (J^d+J^{DC}_{shift})
\end{equation}
Substituting the form for the diffusion current transforms this equation into a diffusion equation given by
\begin{equation}\label{eq:diffeqn}
	\partial_ t \rho = D \partial ^2_x \rho + f(x)
\end{equation}
where the source term is given by the gradient of shift current and has the form:
\begin{equation}\label{eq:sourceterm}
	f(x) = -\partial_x J^{DC}_{shift} = 2\alpha A_0(x) \partial_x A_0(x)
\end{equation}
Since there is no charge density initially (i.e. $\rho(x,0) = 0$), the solution of Eq. \ref{eq:diffeqn} is given by
\begin{equation}\label{rhosoln}
	\begin{split}
		& \rho (x,t) = \int _{-\infty}^{\infty} dx' \, f(x') \int _{0}^t d\tau \, \frac{e^{-\frac{(x-x')^2}{4D\tau}}}{\sqrt{4\pi D \tau}}
	\end{split}
\end{equation}
The diffusion current can be obtained by differentiating the solution in Eq. \ref{rhosoln} w.r.t. $x$ to be
\begin{equation}\label{eq:expdifftime}
	J^d(x) =  \int _{-\infty}^{\infty} dx' \, f(x') \, \frac{(x-x')}{4 \sqrt{\pi D}} \int_0^t d\tau \, \tau ^{-\frac{3}{2}} e^{-\frac{(x-x')^2}{4D\tau}}
\end{equation}
Performing a change of variable, $p = \frac{|x-x'|}{2\sqrt{D\tau} }$, the limits for $p$ becomes $\infty$ and $\frac{|x-x'|}{2\sqrt{Dt}}$ respectively. In the long time limit set by the illumination length and diffusion coefficient($t \gg \frac{|x-x'|^2}{D}$), we can approximate the lower limit to be zero $\frac{|x-x'|}{2\sqrt{Dt}} \to 0$. In this long time limit, performing the Gaussian integral we obtain the expression for the diffusion current to be
\begin{equation}\label{eq:expdiffcurr}
	J^d(x) = \frac{1}{2}  \int_{-\infty} ^{\infty} dx'\, f(x') \, sgn(x-x')
\end{equation}
Substituting Eq. \ref{eq:sourceterm} in Eq. \ref{eq:expdiffcurr} and integrating by parts we obtain $J^d(x) = -J^{DC}_{shift}(x)$, i.e. the diffusion current cancels the shift current in steady state. A similar situation is discussed in Ref. \cite{Strumanrev1980} where the photocurrent is compensated by the ohmic current in a circuit with open boundary condition. The cancellation of the shift current by the diffusion current relates the shift current to the density profile as:
\begin{equation}\label{eq:shiftdengrad}
	J^{DC}_{shift} = - J^d =  D \, \partial_x \rho
\end{equation}
The charge density $\rho (x)$ can be measured capacitively using the gate shown in Fig. \ref{fig:inhom}. This provides a way to measure the shift current without contacts \cite{Yakovenko}.\\
Using $A_0(x') = \int dx'' \, A_0(x'') \delta (x'-x'')$ we can rewrite the expression for the diffusion current (Eq. \ref{eq:expdiffcurr}) as
\begin{equation}\label{eq:expdiffnonloc}
	J^d(x) = \alpha \int_{-\infty} ^{\infty} dx' \, dx'' \,   A_0(x')A_0(x'') sgn (x-x') \partial_{x'} \delta (x'-x'')
\end{equation}
From this expression, it is clear that the diffusion current in case of inhomogeneous excitation profile (i.e. $A_0(x)$ not constant in space) is non-local. This implies that the resultant DC current from the response, given by Eq. \ref{response}, must contain a spatially non-local contribution.\\
The non-local response is quite unexpected in the light of Sec. \ref{sec:secordoptresp}. This is because the time-ordered response functions $\chi_{n,2}$ and $\chi_{n,3}$ can be related by analytic continuation to imaginary time correlation functions, which in turn can be written as derivatives of an effective action \cite{Sachdev}. This is  similar to finite temperature correlation functions in classical statistical mechanics, where correlation functions decay on the scale of a correlation length \cite{Goldenfeld}. One subtlety to keep in mind here is that $\chi_{n,2}$ and $\chi_{n,3}$ in Eqs. \ref{eq:chin2}-\ref{eq:chin3} also involve time. However, analytic continuation from imaginary time suggests that the time-ordered correlations at frequencies larger than temperature are also expected to be local in space and time on a scale determined by correlation lengths and times \cite{Sachdev}. Thus one does not expect $\chi_{n,2}$ or $\chi_{n,3}$ to contribute to the non-local response in Eq. \ref{eq:expdiffnonloc}.\\
Assuming that $\chi_{n,2}$ and $\chi_{n,3}$ are local for the reasons outlined in the previous paragraph, the only term in Eq. \ref{response} that can incorporate non-locality of $J^d$ is the term involving OTOC, as appears in the expression of $\chi_{otoc}$ in Eq. \ref{eq:chiotoc}. The total current is written as $\langle J \rangle_{DC}=J^{DC}_{shift} + J^d$. Since the shift current response $J^{DC}_{shift}$  is local, we can compare the non-local contribution contained in $\chi_{otoc}$ with the non-local diffusion response  (Eq. \ref{eq:expdifftime}) to write the non-local part of $\chi_{otoc}$ as \small 
\begin{align}\label{eq:otocnonlocal}
\chi_{otoc}(r,x_1,x_2)\simeq  \partial_{x_1} \delta (x_1-x_2) \, \frac{(x-x_1)}{4 \sqrt{\pi D}} \int_0^t d\tau \, \tau ^{-\frac{3}{2}} e^{-\frac{(x-x_1)^2}{4D\tau}}
\end{align}
\normalsize for $|x-x_1|,|x-x_2|\gg \xi_{th}$, where $\xi_{th}$ is a thermal correlation length. Thus, the semiclassical approach can be used to constrain the non-linear response correlator (Eq. \ref{eq:chiotoc}), which cannot be computed from equilibrium field theory methods. In principle, this response can be computed from a Keldysh field theory approach \cite{Larkin1968}, but it is clear that these correlators will be qualitatively more non-local compared to equilibrium correlators.

\subsection{Finite voltage and size effect on BPVE\label{subsec:finsizeeff}}
Finite voltage that may be generated in the presence of spatial variations can have an effect on the electron-hole gas as mentioned at the end of Sec. \ref{subsec:semiclasshomo}. The carrier density of this electron hole plasma, $n$, is limited since the rate of exciton generation, for zero temperature, is balanced by recombination rate. We emphasise here that the time scale for recombination processes to lead to a steady state is inversely proportional to some power of EM amplitude ($\propto \frac{1}{A_0^{\alpha}}, \; \alpha >0$) till which the response function formalism is not applicable. We circumvented this problem in the last sub-section by working at finite temperature. The analysis of the balancing of the recombination process at zero temperature considered in this sub-section, assuming $\alpha=1$ for free diffusion of carriers, leads to a steady state carrier density with $n^2 \propto A_0^2$. The shift current in the absence of a voltage can only be stable for perfect translation invariance with periodic boundary conditions. Any variations of parameters in the system that would change the shift current, leads to a finite voltage. Let us now consider boundary conditions different from periodic so that a finite voltage $V$ can be generated over length $L$ of the material. The voltage $V$ will lead to a combination of drift and diffusion current among the carriers which is given by 
\begin{equation}\label{eq:drift}
J^{d}=\frac{n e^2 \tau _d}{m}\frac{V}{L}\propto \frac{A_0 V}{L}
\end{equation}
where $\tau _d$ is the scattering time for the charge carriers.\\
The combined system with a BPVE together with the drift current can be thought of as a current source $J^{d}$ in parallel with a resistance $R\propto L/(\tau _d A_0)$. The maximum voltage generated by this process is $V_{max}=J^{DC}_{shift} R \propto L A_0/\tau_d $. This voltage can become anomalously large in the limit of large L which is consistent with experimental measurements reported in Ref. \cite{Fridkin2001} and the discussions in Ref. \cite{Strumanrev1980}. The maximum power that can be extracted from this system is 
\begin{equation}\label{eq:power}
P_{max, BPVE}=(J^{DC}_{shift})^2 R\propto \frac{A_0 ^3 L}{\tau_d}.
\end{equation}
Note that total incident power scales as $P_{inc} \propto A_0^2 L$. Thus, despite the high open circuit voltage $V_{max}$, the efficiency scales as $P_{max,BVPE}/P_{inc} \propto A_0$ which is quite small and in agreement to the experimental results as discussed in Ref. \cite{Fridkin2001}.

\section{Quantum approach\label{sec:quantumapproach}}
In this section we go beyond the limit of weak optical excitation strength by using a more quantum mechanical master equation based approach to the photocurrent. The problem of the semiconductor in EM field can be described by Eq. \ref{eq:fullham} where $\hat{H}_0$ is the Bloch Hamiltonian as in Eq. \ref{eq:Bloch}. The EM coupled Hamiltonian (Eq. \ref{eq:fullham}) is time periodic with period $T \equiv \frac{1}{\Omega}$. The time periodic Hamiltonian in Eq. \ref{eq:fullham} can be expressed in Floquet basis as
\begin{equation}\label{floqham}
	\hat{H}_F = \sum_k \left( \epsilon ^{u}_k \;  f^{u \dagger}_kf^u _k + \epsilon ^{d}_k \; f^{d \dagger}_kf^d _k \right)
\end{equation}
where $\epsilon ^{u,d}_k \in [ 0,\Omega ]$ are the two quasi energies and $f^{u,d}_k$ correspond to Floquet state operators (see Appendix \ref{app:floqtheo} for derivation of Floquet spectrum).\\
In the quantum mechanical treatment, we include the phonons explicitly with a Hamiltonian that is written as
\begin{equation}\label{eq:phham}
	\hat{H}_p = \sum_{q,s} \omega ^s_q \, a^{s \dagger }_q a^s_q
\end{equation}
where $a^s_q$ is annihilation operator for phonon of mode $s$ and momentum $q$. In absence of any e-ph coupling the evolution operator for combined e-ph system is given by
\begin{equation}\label{eq:evol}
	\begin{split}
		& \hat{U}_0 \equiv \hat{U}_F \otimes \hat{U}_p = e^{-\iota \hat{H}_F t}  \otimes  e^{-\iota \hat{H}_p t} .
	\end{split}
\end{equation}
  The phonons couple to the electrons through the e-ph interaction term in the Hamiltonian given by
\begin{equation}\label{eq:epcoup}
	\hat{V} = \sum^{k,g,s}_{q=k-g} \left( V_1 \, c^{\dagger}_k c_g +V_2 \, v^{\dagger}_k v_g  \right) \left( a^{s}_{q} + a^{s \dagger}_{-q} \right)
 \end{equation}
 where $V_1$($V_2$) is the amplitude of intra band scatterings within conduction (valence) band.\\
 The equation of motion for e-ph combined density operator $\hat{\rho _I}(t)$ , in the interaction picture basis that is rotating according to Eq. \ref{eq:evol}, is given by 
 \begin{equation}\label{eq:liuoville}
 	\partial _t \hat{\rho}  _I  = -\iota \comm{\hat{V}_I}{\hat{\rho}  _I} .
 \end{equation}
 Note that $\hat{V}_I$ is obtained from Eq. \ref{eq:epcoup} by transforming the field operators $c_k, v_k$ and $a^s_q$ to the interaction picture basis. In a standard procedure (outlined in Appendix \ref{app:stepsrateq}) to obtain the dynamical equation for the reduced electronic density operator, $\hat{\rho} ^e_I$, Eq. \ref{eq:liuoville} is solved up to second order in $|V_{1,2}|$ following which the phonon operators are traced out under  Markovian approximation. The resultant equation of motion for $\hat{\rho} ^e_I$ becomes 
 \begin{equation}\label{eq:dynreddenop}
 	\begin{split}
		 \partial_t \hat{\rho} ^e _I = -Tr \left \{ \comm{\hat{V}_I(t)}{\int _{-\infty}^{t} \, dt' \comm{\hat{V}_I(t')}{\hat{\rho} _I(t) }} \right \}_p .
	\end{split}
 \end{equation}
$\hat{\rho}^e_I$ can be represented more concretely in terms of its matrix elements in Floquet basis:
 \begin{equation}\label{eq:occu}
 	\Pi ^{ab}_k(t) \equiv   Tr  \left \{ \hat{\rho }^e _S(t) \; f^{a\dagger}_k f^b_k \right \}  _{e} \equiv Tr  \left \{ \hat{\rho} ^e _I (t) U^{\dagger}_F \hat{\Theta} ^{ab}_k U_F \right \} _{e}
 \end{equation}
where in the last line we have used the cyclic property of trace.
\subsection{Equation for equilibrium distribution\label{subsec:rateq}}
To obtain an equation of motion for $\Pi^{ab}_k$ we differentiate Eq. \ref{eq:occu} and use Eq. \ref{eq:dynreddenop} in the similar way described in \cite{Tom2015}. In the process we assume zero temperature for the phonon bath and translational invariance for the electronic system (see Appendix \ref{app:stepsrateq}) to obtain the final form of the equation of motion for $\Pi^{ab}_k$ to be\small
 \begin{equation}\label{eq:rateq}
    		\begin{split}
            	&\partial _t \Pi _k =- \iota \comm{\Pi_k}{F_k} -\Pi_k \, A_k  -  A^{\dagger}_k \, \Pi _k+(\hat{I}-\Pi _k) D_k + D^{\dagger}_k (\hat{I}-\Pi _k)
    		\end{split}
\end{equation}
 \normalsize  where $F_k$ can be thought of re-normalized Hamiltonian (local in $k$) and is given by,
 \begin{equation}\label{eq:reham}
 	\begin{split}
     	& F^{a b}_k =\frac{\epsilon ^u_k - \epsilon ^d_k}{2}\delta ^{ab} +\iota \, V^{b a}_{kk}\sum ^{\beta \rho}_{g} (\tilde{V}^{\rho \beta}_{gg}-(\tilde{V}^{\beta \rho}_{gg})^*) \Pi ^{\rho \beta}_g\\        
        \end{split}
 \end{equation}
 Here the Fourier components of the matrix elements of $V$ and $\tilde{V}$ are given by,
        \begin{equation}\label{eq:velements}
        	\begin{split}
        		&   V^{\alpha \beta}_{kg}(n) =\sum_m \langle \psi ^{\alpha}_k(m+n) | \hat{V} ^S| \psi ^{\beta}_g (m) \rangle , \\
                &  \tilde{V}^{\alpha \beta}_{kg}(n) =G V^{\alpha \beta}_{kg}(n) \, \Theta \left( \epsilon ^{\alpha}_k - \epsilon ^{\beta} _g + n\Omega \right).
        	\end{split}
        \end{equation}
 where $|\psi ^{\alpha}_k (n)\rangle$ is the nth frequency component of the Floquet state $|\psi ^{\alpha}_k(t)\rangle$ and $\hat{V}^S = \sum _{p,p'} \left( V_1 \, c^{\dagger}_p c_{p'} +V_2 \, v^{\dagger}_p v_{p'} \right)$. $G$ is the constant density of states for the phonon bath and  $\Theta (x) $ is the Heaviside step function.\\        
        The structure of the tensor $D_k$,
        \begin{equation}\label{eq:Dtensor}
        	D^{a b}_k \equiv \sum ^{\beta \rho}_{g}  V^{b \beta}_{kg} \tilde{V}^{\rho a}_{gk} \, \Pi ^{\rho \beta}_g 
        \end{equation}
in Eq. \ref{eq:rateq}, can be understood by considering the diagonal limit of $\Pi_k$. In this limit the product $\Pi ^{\beta \beta} _g (1-\Pi ^{\alpha \alpha}_k)$ along with the corresponding matrix elements represent the probability of scattering from $|\psi ^{\beta}_g \rangle$ to $|\psi ^{\alpha}_k \rangle$ state. The $\Theta$ function ensures the initial state has higher quasi energy than the final state thus only allowing processes that dissipate energy to the phonon bath. Similarly the tensor $A_k$ is involved in reverse scattering processes (for instance from $|\psi^{\alpha}_k \rangle$ to $|\psi ^{\beta}_g\rangle$ scatterings) and is given by,
  \begin{equation}\label{eq:Atensor}
       	\begin{split}
       		A^{a b}_k \equiv \sum ^{\beta \rho}_{g}   V^{b\beta}_{kg} (\tilde{V}^{a \rho }_{kg})^* (\delta ^{\beta \rho} - \Pi ^{\rho \beta}_g).
       	\end{split}
  \end{equation}
       The steady state version of Eq. \ref{eq:rateq} can be obtained by setting $\partial _t \Pi ^{ab}_k=0$. The result is a set of non linear algebraic equations involving $\Pi ^{\alpha \beta} _k \Pi ^{\gamma \delta} _g$ terms and thus finding exact, simultaneous solutions for $\left \{ \Pi ^{\alpha \beta}_k \right \}$ is numerically intensive.\\
        We note since the EM amplitude is much smaller than the bandwidth of the electronic energy spectrum ($A_0|J^{\perp}_{0,k}| \ll |\Delta (0)- \Delta (\pi)|$), $\Pi ^{\alpha \beta}_k$ differs from its ground state value only in a small region (degeneracy region) of $O[A_0]$ in Brillouin zone. Hence scattering contributions between two degeneracy regions (being smaller by order of $A_0$) can be ignored.\\
       Since $\Pi^{\alpha \beta}_k$ is close to its ground state value for almost all of the Brillouin zone, we can use the ground state values for $\Pi^{\alpha \beta}_g$ in the tensors $A ^{\alpha \beta}_k, D^{\alpha \beta}_k, F^{\alpha \beta}_k$ to linearize the steady state equation of motion which can be readily solved analytically. The linearized equation of motion, for the $k$ points inside the degeneracy region, is obtained to be (see Appendix \ref{app:stepslin} for details)
        \begin{equation}\label{eq:linrateq}
        	\begin{split}
		& \iota \comm{ \tilde{\epsilon} _k}{\Pi_k} = -\acomm{M_k}{\Pi _k} + \Lambda_k 
		   \end{split}
       \end{equation}
where $\tilde{\epsilon}^{ab}_k = \frac{1}{2}(\epsilon ^a_k - \epsilon ^b_k)\delta ^{ab}$ is the Hamiltonian in Floquet basis. $M_k$ and $\Lambda_k$ can be deduced from $A_k$ and $D_k$ as defined in Eq. \ref{eq:Dtensor}-\ref{eq:Atensor}, to be 
 \begin{equation}\label{eq:melements}
 	\begin{split}
		& M^{\alpha \beta}_k \simeq  q  \sum_{n<0} \langle \psi^{\alpha}_k (n)|\hat{S}_v|\psi^{\beta}_k(n)\rangle + p \sum_{n\geq 0} \langle \psi^{\alpha}_k(n)|\hat{S}_c|\psi^{\beta}_k(n)\rangle, \\
        &\Lambda ^{\alpha \beta}_k \simeq 2  q  \sum_{n<0} \langle \psi^{\alpha}_k (n)|\hat{S}_v|\psi^{\beta}_k(n)\rangle.
	\end{split}
 \end{equation}
 Here $p$ and $q$ are effective scattering rates for electrons and holes defined as:
 \begin{equation}\label{eq:scattrates}
 	p \equiv f_> |V_1|^2G \,\,\, \text{and} \,\,\, q \equiv f_> |V_2|^2G
 \end{equation}
 where $f_>$ is the fraction of the Brillouin zone such that $\Omega > 2\Delta _k$. $\hat{S}_v$ and $\hat{S}_c$ are the projector operators on the valence and conduction band subspaces respectively defined as
 \begin{equation}\label{eq:projectors}
 	\hat{S}_c \equiv \sum _k c^{\dagger}_k c_k, \,\,\,\,\,\,  \hat{S}_v \equiv \sum _k v^{\dagger}_k v_k .
 \end{equation}
 The solutions of Eq. \ref{eq:linrateq} can be used to calculate current responses.
 \subsection{Expression for current response\label{subsec:jdc}}
 The matrix elements of current operator in the basis of Floquet states are
\begin{equation}\label{eq:currmatelements}
	\begin{split}
		& C ^{\alpha \beta}_k \equiv \langle \psi ^{\alpha}_k | \hat{J}(t) | \psi ^{\beta}_k \rangle .
	\end{split}
\end{equation}
 Solving the linear equations in Eq. \ref{eq:linrateq} and calculating $C^{\alpha \beta}_k$ according to Eq. \ref{eq:currmatelements} (Appendix \ref{app:curr} contains expressions for the steady state matrix elements, $\Pi ^{\alpha \beta}_k$ and $C^{\alpha \beta}_k$), we obtain expression for the current response. The AC current can be obtained by computing the $\pm \Omega$ frequency components of $C^{\alpha \beta}_k$ and combining with the steady state solution for $\Pi ^{\alpha \beta}_k$ (see Appendix \ref{app:curr}). The resultant response for the AC current up to leading order in electric field amplitude is
\begin{equation}\label{eq:jac}
	\begin{split}
		& J^{AC}_k(\Omega) = A_0|J^{\perp}_{0,k}|^2 L_k \left[ 1-\iota+\frac{p+q}{\epsilon _k} +\frac{J^z_{1,k}(\Omega - 2\Delta_k)}{|J^{\perp}_{0,k}|^2} \right]\\
    & \text{with} \;\;  L_k \equiv \frac{\left( \frac{\Omega}{2}-\Delta _k\right)}{2\epsilon _k} \frac{2q+t_k (p-q)}{(p+q)^2+(2\epsilon _k)^2}
	\end{split}
\end{equation}
where $t_k$ is given by
\begin{equation}\label{eq:trace}
	 t_k \equiv Tr \{ \Pi _k\} = \frac{1 - \frac{2(p-q)}{p}\frac{A_0^2|J^{\perp}_{0,k}|^2}{(p+q)^2+(2\epsilon _k)^2}}{1 + \frac{(p-q)^2}{pq}\frac{A_0^2|J^{\perp}_{0,k}|^2}{(p+q)^2+(2\epsilon _k)^2}} .
\end{equation}
Similarly, computing the zero frequency component of $C^{\alpha \beta}_k$, we find the DC current up to second order in electric field magnitude to be \small
\begin{equation}\label{eq:jdc}
	J^{DC}_{shift}(k) = Tr \{ \Pi _k C_k \}_e= 2A_0 ^2 \left( \vec{J}_{0,k} \cross \vec{J}_{1,k} \right) _z \frac{2q + t_k (p-q)}{(p+q)^2 + (2\epsilon _k)^2} .
\end{equation}
\normalsize The total DC current is obtained integrating Eq. \ref{eq:jdc} over Brillouin zone as
\begin{equation}\label{eq:totcurr}
	J_{shift}^{DC} = \frac{1}{2 \pi}\int _{BZ}dk \, J^{DC}(k).
\end{equation}
If the EM perturbation is the smallest parameter (i.e. $A_0|J^{\perp}_{0,k}|\ll p+q$), from Eq. \ref{eq:trace} $t_k \simeq 1$ which implies 
\begin{equation}\label{eq:jdclim}
	\lim _{A_0\ll (p+q)} J^{DC}_{shift}(k) =2 A_0^2 \left( \vec{J}_{0,k} \cross \vec{J}_{1,k}\right)_z \frac{p+q}{(p+q)^2+(2\epsilon _k )^2} .
\end{equation}
Recognizing the total scattering rate as $\eta = p+q$ and using the limit, $\lim _{A_0 \to 0} \epsilon _k = \frac{\Omega}{2}-\Delta _k$, we recover the semi classical result obtained in Eq. \ref{eq:DCsclim}.\\
$J^{DC}_{shift}(k)$ can also be evaluated in the opposite limit of weak scattering (i.e. $p,q\ll A_0|J^{\perp}_{0,k}|$) to obtain
\begin{equation}\label{eq:jdclim2}
	\begin{split}
		& \lim_{A_0|J^{\perp}_{0,k}| \to 0} \; \; \lim_{p,q \ll A_0|J^{\perp}_{0,k}|} J^{DC}_{shift}(k) \\
        & =\frac{A_0 \left( \vec{J}_{0,k} \cross \vec{J}_{1,k} \right) _z}{2|J^{\perp}_{0,k}|} (2q+t_k (p-q)) \delta \left(\frac{\Omega}{2}-\Delta_k \right) .
	\end{split}
\end{equation}
Putting this expression into Eq. \ref{eq:totcurr} it is easy to see that the resultant DC current scales linearly with the electric field in this limit, i.e. $J^{DC}_{shift} \propto A_0$. This result contradicts Kubo formula by producing a linear scaling with the external oscillating EM field for the DC current.
\section{Numerical evaluation of the weak damping limit\label{sec:numerics}}
According to the analytic results in Eq. \ref{eq:jdclim} and Eq. \ref{eq:jdclim2}, the DC current response scales quadratically at small $A_0$ and linearly at large $A_0$ (smaller and larger than e-ph scattering strengths). To study how the crossover occurs between the two regimes, we perform the integral in Eq. \ref{eq:totcurr} using the expression for $J^{DC}_{shift}(k)$ given by Eq. \ref{eq:jdc} for the Rice-Mele model \cite{Rice1982}. We also numerically study the scaling property of AC current magnitude for the same model. The Rice-Mele model is described by the following tight binding Hamiltonian \cite{Rice1982}
\begin{equation}\label{eq:ricemeleham}
	\begin{split}
		& \hat{H}_0 = \sum_n \left( h \, a^{\dagger}_n b_n + h' \, a^{\dagger}_n b_{n-1} + h.c. \right) + d \left( a^{\dagger}_n a_n - b^{\dagger}_n b_n \right) \\
        & = \sum_k \Delta _k \left( c^{\dagger}_k c_k - v^{\dagger}_k v_k \right) \\
        & \mathrm{with} \,\,\,\, \Delta _k \equiv \sqrt{h^2 + h'^2 + 2hh' \cos (k) +d^2} .
	\end{split}
\end{equation}
\vspace{.1in}
\begin{tikzpicture}\label{pic:latt}
    	\draw (3,0) -- (9,0);
       
        \foreach \x in {5,7,9}
        \draw (\x ,.05) -- (\x - 1, .05);
        
        \foreach \x in {3,5,7,9}
   \draw (\x cm,1pt) -- (\x cm,-1pt) node[anchor=north] {A};
   \foreach \x in {4,6,8}
   \draw (\x cm,1pt) -- (\x cm,-1pt) node[anchor=north] {B};
   \foreach \x in {3,5,7,9}
   \fill[black] (\x cm,0) circle (.12cm);
   \foreach \x in {4,6,8}
   \fill[blue] (\x cm,0) circle (.12cm);
   \foreach \x in {4,6,8}
   \draw (\x -.5,1pt) -- (\x -.5,-1pt) node[anchor=south] {t};
   \foreach \x in {5,7,9}
   \draw (\x -.5,1pt) -- (\x -.5,-1pt) node[anchor=south] {t'};
        
    \end{tikzpicture}
    \vspace{.1in}
    
The EM vector potential $A(t)$ introduces phases to the hoping terms according to Peierls substitution as
\begin{equation}\label{eq:peierlssub}
	a^{\dagger}_n b_n \rightarrow e^{-\iota \frac{A(t)}{2}} \, a^{\dagger}_n b_n, \,\,\,\, a^{\dagger}_n b_{n-1} \rightarrow  e^{\iota \frac{A(t)}{2}} \, a^{\dagger}_n b_{n-1} .
\end{equation}
Transforming to Fourier space, the perturbed Hamiltonian, in the atomic ($a_k, b_k$) basis,  becomes
\begin{equation}\label{eq:emcouplrmham}
	\begin{split}
		& \hat{H} = \sum_k \left(  \begin{array}{cc}
		a^{\dagger}_k & b^{\dagger}_k
		\end{array} \right) \vec{H}_k \cdot \vec{\sigma} \left( \begin{array}{c}
		a_k \\
        b_k
		\end{array} \right)\\
        & \text{with} \\
        & \vec{H}_k =-\left[( h  + h') \cos \frac{k-A}{2},(-h+ h') \sin \frac{k-A}{2},-d \right] .
	\end{split}
\end{equation}
In terms of system parameters, $\left( \vec{J}_{0,k} \cross \vec{J}_{1,k} \right) _z = -\frac{d}{8\Delta_k } \left( h'^2 - h^2 \right)$ where $\vec{J}_{0,k} \equiv \partial _A \vec{H}_k(A)|_{A=0}$ and $\vec{J}_{1,k} \equiv \partial ^2_A \vec{H}_k (A)|_{A=0}$.\\
\begin{figure}[t]
	\vspace{.2in}
	\centering
    \includegraphics[width=7cm]{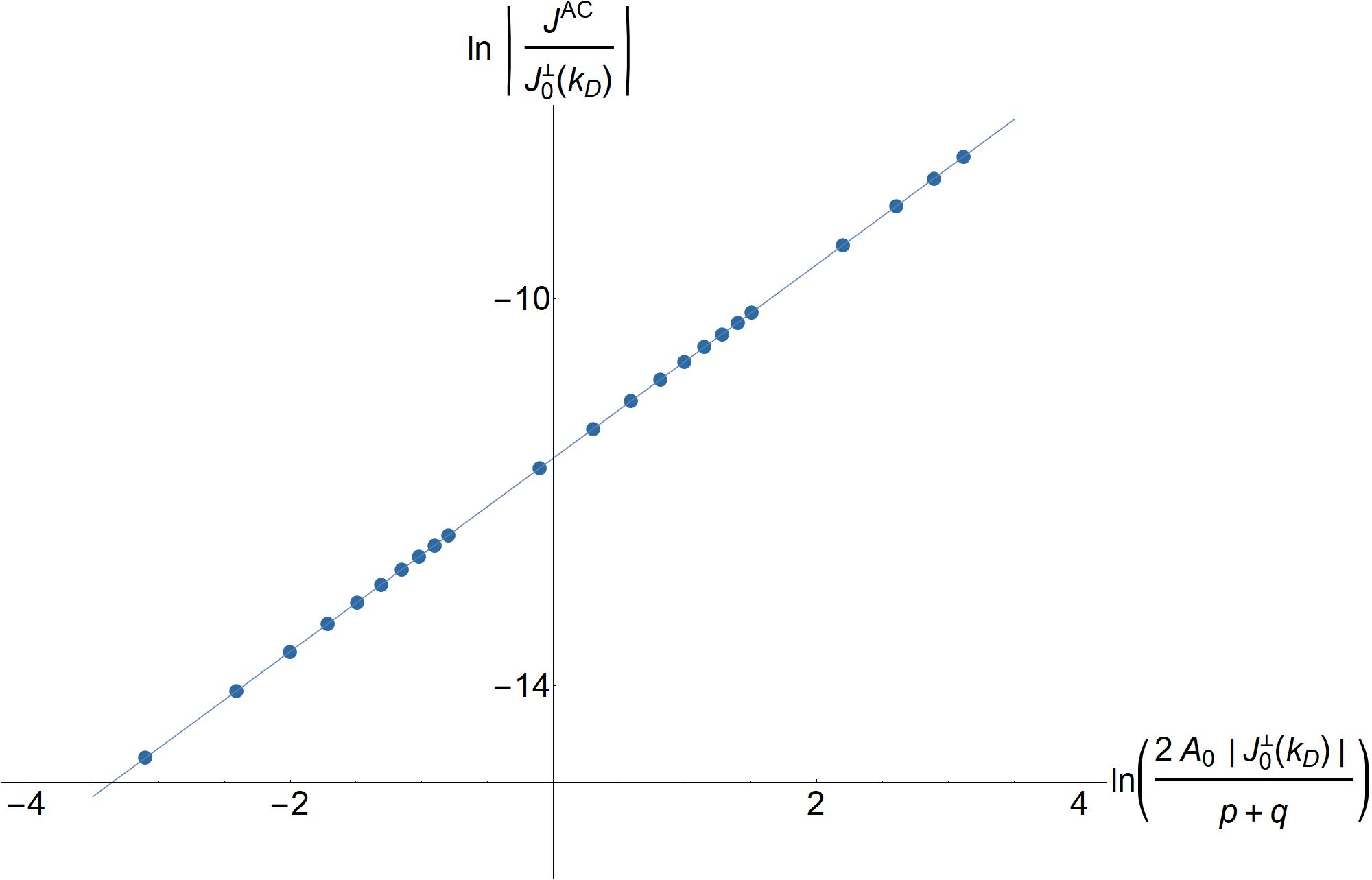}
   \caption{  $ln \left| J^{AC}/|J^{\perp}_{0,k_D}| \right|$ vs $ln \left|2A_0|J^{\perp}_{0,k_D}|/(p+q)\right|$ plot. The equation for the linear fit is $y=-11.65+x$.}
   \label{fig:linresp}
\end{figure}
\begin{figure}[t]
	\vspace{.2in}
	\centering
    \includegraphics[width=7cm]{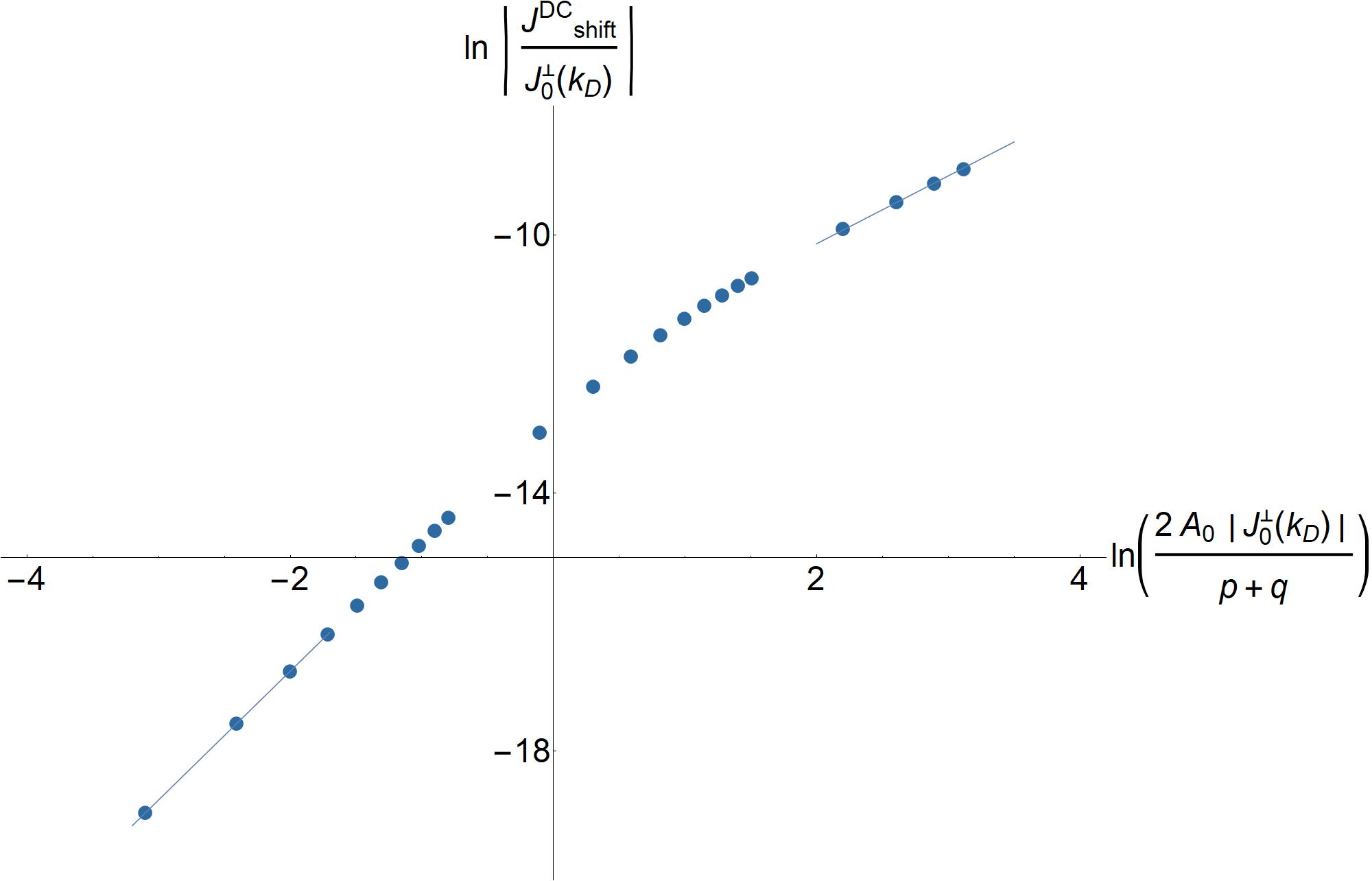}
   \caption{  $ln \left| J^{DC}_{shift}/|J^{\perp}_{0,k_D}| \right|$ vs $ln \left|2A_0|J^{\perp}_{0,k_D}|/(p+q)\right|$  plot. Linear fits in regimes $A_0|J^{\perp}_{0,k_D}| \ll p,q$ and $A_0|J^{\perp}_{0,k_D}|\gg p,q$ show a switch in slope from $2$ to $1$ as $A_0$ is increased across the e-ph coupling strength. The linear fit equations are $y=-13.3 + 2 x$ and  $y=-12.8 + 1.05 x$ respectively.}
   \label{fig:nonlinresp}
\end{figure}

The AC current for this Rice-Mele model is obtained by numerical integration of the expression given by Eq. \ref{eq:jac} over the Brillouin zone with fixed values for $p,q$. The resulting values for $|J^{AC}|/|J^{\perp}_{0,k_D}|$ are plotted as a function of $2 A_0|J^{\perp}_{0,k_D}| /(p+q)$ on a log scale in Fig. \ref{fig:linresp} where $k_D$ is the momentum of the degeneracy point (with the property, $ 2\Delta _{k_D} = \Omega$). The points on the plot fit to a straight line with slope 1. Thus the AC current response scales linearly in a range of values from $A_0|J^{\perp}_{0,k_D}|\ll (p+q)$ to $A_0|J^{\perp}_{0,k_D}|\gg p+q$. This scaling is as expected from conventional linear response theory. On the other hand same plot for the shift current, $|J^{DC}_{shift}|/|J^{\perp}_{0,k_D}|$, shows a crossover of scaling with $A_0$ as shown in Fig. \ref{fig:nonlinresp}. In the limit of weak electric field, $\left( ln \left|2A_0|J^{\perp}_{0,k_D}|/(p+q)\right|<-1 \right)$, the points fit to a straight line of slope 2 whereas in the limit of weak scattering $\left( ln \left|2A_0|J^{\perp}_{0,k_D}|/(p+q)\right|>1 \right)$, the corresponding fit line has slope 1. Therefore, the DC current response indeed shifts scaling from quadratic to linear through a region where the electric field and the scattering strengths are comparable $\left( ln \left|2A_0|J^{\perp}_{0,k_D}|/(p+q)\right| \sim 0 \right) $. Thus the e-ph coupling strength sets an energy scale for the EM field strength for a non-linear to linear crossover of the shift current response.
\section{Conclusion}
We have studied an example of a non-linear response i.e. the BPVE in a system with electron-phonon interactions. Since the response is dominated by resonant transitions, we have limited our treatment to a two band model. As we found in Sec. \ref{sec:secordoptresp}, the BPVE contains OTOC terms that cannot be computed by simple diagrams. However, in Sec. \ref{sec:semiclass}, we find that the BPVE can be understood semiclassically from the dipole moment of generated excitons. The semiclassical limit assumes that the scattering rate is large compared to the driving force. We remedy this in Sec. \ref{sec:quantumapproach} using a master equation approach. From this we find that in the small scattering rate limit the DC current scales linearly with the vector potential instead of quadratically. This signals that this result is beyond the response function formalism in Sec. \ref{sec:secordoptresp}. The AC current doesn't show any such anomalous scaling behavior. \\
The DC current we obtain in the small $A_0$ limit is consistent with the non-interacting result for BPVE obtained previously. The breakdown seen in Sec. \ref{sec:quantumapproach} at larger $A_0$ shows that electron-phonon interaction affects the result in an implicit way. However, what is interesting from comparing to the general response in Sec. \ref{sec:secordoptresp} is that only the $\chi_{n,2}$ term contribute to the shift current in configurations with homogeneous excitation profiles. This term is really conventional linear response of $\hat{J}_0$ to $\hat{J}_1$ contributions in the BPVE and can be computed from equilibrium field theory. However, as shown in Sec. \ref{subsec:inhomexcitation}, the situation is quite different for inhomogeneous excitation profiles. In this case, owing to inhomogeneous carrier density generated by the incident light an additional diffusion current response is generated. This modulation of the density profile provides an alternative way to measure the shift current capacitively without the use of contact leads. The diffusion current contribution is not captured by the simple shift current that can be computed from $\chi_{n,2}$ and requires the computation of $\chi_{otoc}$. As discussed in Sec. \ref{sec:secordoptresp}, since $\chi_{otoc}$ is not a time-ordered correlator, it cannot be computed from conventional equilibrium field theory but rather requires a Keldysh technique. In this work, we have circumvented this using a semi-classical approach that is of limited validity in the context of weak interactions to place a constraint on the long ranged part of $\chi_{otoc}$ (Eq. \ref{eq:otocnonlocal}). The full quantum mechanical computation of $\chi_{otoc}$ using Keldysh field theory for strongly interacting systems where the semiclassical estimate would break down, would be an interesting future direction. \\
We acknowledge interesting comments from Victor Yakovenko at the inception of this project. This work has been supported by the National Science Foundation NSF DMR1555135 (CAREER)  and JQI-NSF-PFC (supported by NSF grant PHY-1607611). Jay Sau is grateful to the Aspen Center for Physics where part of the work was completed.

\bibliographystyle{unsrt}
\bibliography{ref}

\begin{appendices}

\section{Derivation of second order response\label{app:response}}
We will derive the current response, $\langle \hat{J}(r) \rangle$, up to second order in the excitation amplitude, $A_0(x)$, due to the EM perturbation, $\int dx \, A(r) \hat{J}(r)$, as given by Hamiltonian in Eq. \ref{eq:fullham}. The space-time dependence of $A(r)$ is defined in Sec. \ref{sec:secordoptresp}. The response can be written as the trace of current operator with the density matrix, as
\begin{equation}\label{eq:respexpval}
	\langle \hat{J}(r) \rangle = Tr \left \{ \tilde{\rho }(t) \mathscr{J}(r) \right \} 
\end{equation}
where $\mathscr{J}(r), \; \tilde{\rho}(t)$ both are in interaction picture, i.e. evolved by the unperturbed Hamiltonian for the system $\hat{H}_0$ as in Eq. \ref{eq:fullham}. The evolution of $\tilde{\rho}(t)$ is thus given by:
\begin{equation}\label{eq:resptimevolrho}
	\tilde{\rho}(t) = \tilde{U}(t) \, \tilde{\rho}_{-\infty} \, \tilde{U}^{\dagger}(t)
\end{equation}
where $\tilde{\rho}_{-\infty}$ is initial equilibrium density matrix for the system, before the EM field was turned on. $\tilde{U}(t)$ is the interaction picture time evolution operator which can be expanded up to second order in $A_0(r)$ from the time-ordered exponential as follows:
\begin{equation}\label{eq:expandu}
	\begin{split}
		& \tilde{U}(t) = \mathcal{T} \left \{ e^{-\iota \int _{r'} \, A(r') \mathscr{J}(r')} \right \}  \\
        & \simeq 1 - \iota \int _{-\infty}^t dt' \int dx' \left[ A(r')\mathscr{J}_0(r') + A^2(r')\mathscr{J}_1(r') \right]  \\
        & -\frac{1}{2} \int _{-\infty}^t dt_1 \, dt_2 \int dx_1 \, dx_2 A(r_1)A(r_2) \mathcal{T} \left \{ \mathscr{J}_0(r_1)\mathscr{J}_0(r_2) \right \} 
	\end{split}
\end{equation}
Using this expansion for $\tilde{U}(t)$ in Eq. \ref{eq:respexpval} and using cyclic property of trace, we obtain for $\langle \hat{J}(r) \rangle$, up to second order in $A_0(x)$:
\begin{equation}
	\begin{split}
		& \langle \hat{J}(r) \rangle \simeq \langle \hat{J}(r) \rangle _0 + \int _{-\infty} ^{\infty} dx' \, A_0(x') \, \chi _{lin} (r,x') \\
        & + \int _{-\infty} ^{\infty} dx' \, A_0^2(x') \, \chi_{n,2} (r,x') \\
        & + \int _{-\infty} ^{\infty} dx_1 dx_2 \, A_0(x_1)A_0(x_2) \, \chi_{n,3} (r,x_1,x_2) \\
        & +\int _{-\infty} ^{\infty} dx_1 \,  dx_2 \, A_0(x_1) A_0(x_2) \, \chi _{otoc} (r,x_1, x_2) 
	\end{split}
\end{equation}
where the linear response function, $\chi_{lin}$ is defined as
\begin{equation}
	\chi _{lin} (r,x') = \iota \int_{-\infty}^t dt' \, e^{\eta t'}\alpha (t') \left  \langle \comm{\mathscr{J}_0(r')}{\mathscr{J}_0(r)} \right \rangle _0 
\end{equation}
The non-linear response function $\chi_{n,2}$ is given by
\begin{equation}
	\chi_{n,2}(r,x') = \iota \int_{-\infty}^t dt' \, e^{2\eta t'} \alpha ^2(t')\left \langle \comm{\mathscr{J}_1(r')}{\mathscr{J}_0(r)} \right \rangle _0
\end{equation}
$\chi_{n,3}$ contains equilibrium time-ordered correlators of three current operators and is given by \small
\begin{equation}
	\begin{split}
		& \chi_{n,3}(r,x_1,x_2) \\
        & = -\frac{1}{2}\int_{-\infty}^t dt_1 \, dt_2 e^{\eta (t_1+t_2)} \alpha (t_1) \alpha (t_2) \left  \langle \acomm{\mathcal{T} \{ \mathscr{J}_0(r_1)\mathscr{J}_0(r_2) \}}{\mathscr{J}_0(r)} \right \rangle _0
	\end{split}
\end{equation}
\normalsize $\chi_{otoc}$ contains out of time ordered correlator as  \small
\begin{equation}
	\begin{split}
		& \chi_{otoc}(r,x_1,x_2) \\
        & =\int_{-\infty}^t dt_1 \, dt_2 e^{\eta (t_1+t_2)} \alpha (t_1) \alpha (t_2) \left \langle  \mathscr{J}_0(r_1)\mathscr{J}_0(r)\mathscr{J}_0(r_2)  \right \rangle _0
	\end{split}
\end{equation}
\normalsize $\langle .. \rangle _0 = Tr \left \{ \tilde{\rho}_{-\infty} .. \right \}$ refers to the ground state expectation value. Since there's no current in equilibrium, the first term vanishes (i.e. no current response in absence of any excitation) and one obtains Eq. \ref{response}. 
\section{Derivation of shift vector\label{app:berryconnection}}
In this appendix we establish the agreement of our result for the DC current as obtained in Eq. \ref{eq:DCsclim} to the conventional form of the same \cite{Sipe2000,Kraut1981,Morimoto2016} as given by,
\begin{equation}
	J^{DC}_{shift} = A_0^2 \int _{BZ} dk \, |u^{cv}_k|^2 \left( \partial_k \phi_k + \alpha ^{cc}_k -\alpha ^{vv}_k \right)
\end{equation}
where $u^{cv}_k = \langle c| \hat{u}_k|v\rangle$ and $\hat{u}_k$ is the velocity operator defined as, $\hat{u}_k =\partial_k \hat{H}_{0,k}$. $\alpha ^{cc}_k$ ($\alpha ^{vv}_k$) is the Berry connection for the $|c_k\rangle$ ($|v_k\rangle$) state and $\phi_k = -\text{Im} [\log u^{cv}_k]$ .
Recalling our result, Eq. \ref{eq:DCsclim}:
\begin{equation*}
	\lim _{\eta \to 0} J_{shift}^{DC} = A_0 ^2 \int _{BZ} \, dk \, ( \vec{J}_{0,k} \cross \vec{J}_{1,k} )_z \delta (\Omega - 2 \Delta _k) ,
\end{equation*}
we recognize that we need to show 
\begin{equation}\label{eq:toprovshiftvec}
	\left( \vec{J}_{0,k} \cross \vec{J}_{1,k} \right)_z = |u^{cv}_k|^2 \left( \partial_k \phi_k + \alpha ^{cc}_k -\alpha ^{vv}_k \right)
\end{equation}
First we observe that
\begin{equation}
	\begin{split}
		& \left( \vec{J}_{0,k} \cross \vec{J}_{1,k} \right) _z = \text{Im} \left[ J_{0,k}^{cv} \, J_{1,k}^{vc} \right] = \text{Im} \left[ u_k^{vc} \left( \partial _k \hat{u}_k\right) ^{cv}  \right]
	\end{split}
\end{equation}
where from the definition of current operators, we have used $\hat{J}_{0,k} = -\hat{u}_k$ and $\hat{J}_{1,k} = \partial _k \hat{u}_k$. Using chain rule for derivatives, we expand the off-diagonal matrix element of $\partial _k \hat{u}_k$ as:
\begin{equation}\label{eq:offdiag}
	\begin{split}
		& \left( \partial _k \hat{u}_k \right) ^{cv} = \langle c|\partial _k \hat{u}_k |v\rangle \\
        & = \partial_k u^{cv}_k - \langle \partial_k c | \hat{u}_k |v\rangle - \langle c | \hat{u}_k | \partial _k v\rangle
	\end{split}
\end{equation}
$ \langle \partial_k c | \hat{u}_k |v\rangle$ can be expanded as:
\begin{equation}\label{eq:expandj1}
	\begin{split}
		&  \langle \partial_k c | \hat{u}_k |v\rangle = \langle \partial_k c | \left[ \partial_k \left( \hat{H}_{0,k}|v\rangle \right) - \hat{H}_{0,k} |\partial _k v\rangle  \right] \\
        & = \partial_k \epsilon ^v_k \langle \partial _k c|v\rangle + \epsilon ^v_k \langle \partial_k c| \partial_k v\rangle - \langle \partial_k c|\hat{H}_{0,k}|\partial_k v\rangle
	\end{split}
\end{equation}
Now inserting Identity operator, $\hat{I}= |c\rangle \langle c| + |v\rangle \langle v|$ in between for the second and third terms in Eq. \ref{eq:expandj1} we obtain
\begin{equation}\label{eq:res1}
	\begin{split}
		& \langle \partial_k c| \hat{u}_k | v\rangle = -a^{cv}_k \left( u^{vv}_k - a^{cc}_k (\epsilon ^c_k - \epsilon ^v _k) \right)
	\end{split}
\end{equation}
where $a^{\alpha \beta}_k= \langle \alpha |\partial_k \beta\rangle$, $u^{vv}_k = \partial_k \epsilon ^v_k$. Interchanging $c$ and $v$ in Eq. \ref{eq:res1} we obtain
\begin{equation}\label{eq:res2}
	\begin{split}
		& \langle \partial_k v| \hat{u}_k | c\rangle = -a^{vc}_k \left( u^{cc}_k - a^{vv}_k (\epsilon ^v_k - \epsilon ^c _k) \right)
	\end{split}
\end{equation}
We observe that the Berry phases are related to the $a^{\alpha \beta}_k$ as $\alpha ^{\alpha \beta}_k = -\iota a^{\alpha \beta}_k$. The off-diagonal term $a^{cv}_k$ can be related to the velocity matrix element as
\begin{equation}
	\begin{split}
		 a^{cv}_k = \langle c|\partial _k \left( \frac{\hat{H}_{0,k}}{\epsilon ^v_k} |v\rangle \right)=-\frac{u^{cv}_k}{\epsilon ^c_k - \epsilon ^v_k}
	\end{split}
\end{equation}
Using complex conjugate of Eq. \ref{eq:res2}in Eq. \ref{eq:offdiag} we obtain:
\begin{equation}
	\left( \partial_k \hat{u} \right)^{cv}_k = \partial_k u^{cv}_k + u_k^{cv} \left[ \frac{u^{cc}_k - u^{vv}_k}{\epsilon ^c_k - \epsilon ^v_k} + a^{cc}_k - a^{vv}_k\right]
\end{equation}
Writing $u^{cv}_k = |u^{cv}_k|e^{-\iota \phi_k}$ we obtain for the imaginary part, $\text{Im} \left[ u^{vc}_k \left( \partial_k\hat{u}_k \right)^{cv} \right]$:
\begin{equation}
	\text{Im} \left[ u^{vc}_k \left( \partial_k\hat{u}_k \right)^{cv} \right] = |u^{cv}_k|^2 \left( \partial_k \phi _k + \alpha ^{cc}_k - \alpha ^{vv}_k \right) = |u^{cv}_k|^2R^{cv}_k
\end{equation}
where the last equality is the connection of the shift vector to the berry connection as defined in previous literature \cite{Sipe2000,Kraut1981,Morimoto2016}.
\section{Floquet Theory for two band system}\label{app:floqtheo}
For a system with time periodic Hamiltonian of frequency $\Omega$, the solutions for the time dependent Schrodinger equation (TDSE) have the following form
\begin{equation}\label{eq:floqsolform}
	|\psi ^{\alpha} _k(t) \rangle=e^{-\iota \epsilon ^{\alpha}_kt} |\phi ^{\alpha}_k(t)\rangle  = e^{-\iota \epsilon ^{\alpha}_kt} \sum_{n} |\psi ^{\alpha}_k(n)\rangle e^{\iota n\Omega t} .
\end{equation}
where the quasi energy $\epsilon ^{\alpha}_k$ can be chosen to belong in a $[0,\Omega]$ interval and $|\phi^{\alpha}_k(t)\rangle$ is time periodic with period $\frac{2\pi}{\Omega}$.\\
Substituting the form of Eq. \ref{eq:floqsolform} in TDSE, one obtains the Floquet equations in frequency space as
\begin{equation}\label{eq:floqeqns}
\sum_n \left( \epsilon ^{\alpha}_k-n\Omega \right) |\psi ^{\alpha}_k(n)\rangle = \sum_m \hat{H}_{k}(m-n) |\psi ^{\alpha}_k(m)\rangle
\end{equation}
where $\hat{H}^k_{m-n}$ corresponds to the $(m-n)\Omega$ frequency component of the Hamiltonian, $\hat{H}_k$. The eigenvalue solutions for Eq. \ref{eq:floqeqns} produce $\epsilon ^{\alpha}_k$ and $|\psi ^{\alpha}_k(n)\rangle$.\\
For the two band case discussed in the main text, the EM perturbation in frequency space can be written as
\begin{equation}\label{eq:pertfloq}
	\begin{split}
		& \hat{H}_p = A_0 \sum _n  J^z_{0,k} (c^{\dagger}_{k,n} c_{k,n \pm 1}-  v^{\dagger}_{k,n}v_{k,n \pm 1}) \\
    & +A_0 \left[ (J^x_{0,k}-\iota J^y_{0,k})c^{\dagger}_{k,n} v_{k,n \pm 1}+h.c. \right] 
	\end{split}
\end{equation}
where the subscript $n$ correspond to $n\Omega$ frequency component for the $c_k$, $v_k$ operators .\\
For the unperturbed Hamiltonian (Eq. \ref{eq:Bloch}) the quasi energies are simply
\begin{equation}
	\epsilon ^{u,(0)}_k = \frac{\Omega}{2}+ \left| \frac{\Omega}{2}-\Delta _k \right| \;\; \text{and} \;\; \epsilon ^{d(0)}_k = \frac{\Omega}{2}- \left| \frac{\Omega}{2}-\Delta _k \right|
\end{equation}
with the corresponding states as
\begin{equation}
	\begin{split}
		& |\psi ^{u(0)}_k \rangle = \Theta (\Omega - 2\Delta _k) |c_k (0)\rangle + \Theta (2\Delta _k - \Omega) |v_k (-1)\rangle ,\\
    	& |\psi ^{d(0)}_k \rangle = \Theta (\Omega - 2\Delta _k) |v_k (-1)\rangle + \Theta (2\Delta _k - \Omega) |c_k (0)\rangle .
	\end{split}
\end{equation}
If $\Omega = 2\Delta _k$, the quasi energies are degenerate and hence to obtain the effect of perturbation (Eq. \ref{eq:pertfloq}) an exact diagonalisation needs to be performed in the vicinity of those degeneracy points.\\
Up to leading order in $A_0$, the shifts can be obtained by diagonalizing the following matrix given by
\begin{equation}
	\begin{split}
		\tilde{H}_k = \left(  \begin{array}{cc}
	\Delta _k & A_0( J^x_{0,k} -\iota J^y_{0,k}) \\
   A_0( J^x_{0,k} +\iota J^y_{0,k}) & \Omega - \Delta _k 
	\end{array} \right)
	\end{split}
\end{equation}
The corresponding eigenvectors can be used to obtain the Floquet state solutions in time domain to be
\begin{equation}\label{eq:floqstates}
	\begin{split}
		& |\psi ^{u}_k(t)\rangle = \frac{e^{-\iota \epsilon ^{u}_kt}}{\sqrt{1+|x_k|^2}} \left( |c_k\rangle+x_k \, e^{-\iota \Omega t}|v_k \rangle \right) ,\\
        & |\psi ^{d}_k(t)\rangle=\frac{e^{-\iota \epsilon ^d_k t}}{\sqrt{1+|x_k|^2}} \left(x^*_k \, |c_k \rangle-e^{-\iota \Omega t}|v_k \rangle \right) \\
        & \mathrm{with} \,\,\,\, x_k = \frac{\frac{\Omega}{2} - \Delta _k +\epsilon _k}{A_0 |J^{\perp}_{0,k}|} \frac{J^x_{0,k} + \iota J^y_{0,k}}{|J^{\perp}_{0,k}|} \\
        & \text{and} \;\; \epsilon ^{u,d}_k = \frac{\Omega}{2} \pm \epsilon _k \;\; \text{with} \;\; \epsilon _k \equiv \sqrt{\left( \frac{\Omega}{2}-\Delta _k \right) ^2 + A_0 ^2 |J^{\perp}_{0,k}|^2} .
	\end{split}
\end{equation}

\section{Derivation of equation of motion for $\Pi^{ab}_k$}\label{app:stepsrateq}
Expanding the solution of Eq. \ref{eq:liuoville} up to second order in $|V_{1,2}|$ we obtain:
 \begin{equation}\label{eq:secordexp}
 	\begin{split}
		&  \hat{ \rho } _I (t) = -\iota \int _{-\infty}^t \, dt' \, \comm{\hat{V}_I(t')}{\hat{ \rho } _I (-\infty )} \\
        &- \int _{-\infty} ^t \, dt' \, \comm{\hat{V}_I (t')}{\int _{-\infty}^{t'} \, dt'' \, \comm{\hat{V}_I(t'')}{\hat{ \rho } _I (t'')}}
	\end{split}
 \end{equation}
 Since, the first term contains single power of bath operators, its trace can be dropped. In that case we note, $|\hat{\rho} _I (t'') - \hat{\rho}_I (t')| \approx O[|V_{1,2}|^2]$ and so up to second order in $|V_{1,2}|$, we can replace $\hat{\rho}_I(t'')$ by $\hat{\rho}_I(t')$ in Eq. \ref{eq:secordexp} to obtain a second order perturbative result for $\hat{\rho}^e_I(t)$ as
 \begin{equation}\label{eq:reddenop}
 	\begin{split}
		& \hat{\rho} _I ^e(t) \equiv  Tr \left \{ \hat{ \rho }^{ep} _I (t) \right \} _p\\
        & \simeq - Tr \left \{ \int _{-\infty} ^t \, dt' \, \comm{\hat{V}_I (t')}{\int _{-\infty}^{t'} \, dt'' \, \comm{\hat{V}_I(t'')}{\hat{ \rho }^{ep} _I (t')}} \right \} _p .
	\end{split}
 \end{equation}
 Differentiating Eq. \ref{eq:reddenop} w.r.t. time one obtains Eq. \ref{eq:dynreddenop}.\\
 Using cyclic property of trace, from Eq. \ref{eq:occu}, $\partial_t \Pi ^{ab}_k$ produces \small
\begin{equation}\label{eq:dynpi}
	\begin{split}
		& \partial _t \Pi ^{ab}_k = -\iota Tr  \left \{ \comm{\hat{\Theta} ^{ab}_k}{\hat{H}_F} \hat{\rho} ^e _S(t) \right \}_e \\
        & -Tr  \left \{ \comm{\comm{\hat{\Theta} ^{ab}_k}{ \hat{U}_0 \hat{V}_I(t) \hat{U}^{\dagger}_0}}{\hat{U}_0 \left( \int _{-\infty}^t dt' \, \hat{V}_I(t') \right) \hat{U}^{\dagger}_0} \hat{\rho}^{ep}_S(t)\right \}_{ep} .
	\end{split}
 \end{equation}
\normalsize
Noting that $U_F (t) \; f^{\alpha \dagger}_k f^{\beta}_g \; U^{\dagger}_F (t) = e^{\iota (\epsilon ^{\alpha}_k - \epsilon ^{\beta}_g)t} \; f^{\alpha \dagger}_k f^{\beta}_g$ and $U_p(t) \; a^{s}_q \; U^{\dagger}_p(t) = a^{s}_q \; e^{\iota \omega ^s_qt}$, we obtain
\begin{equation}
	\begin{split}
		U_0 \, \hat{V}_I(t) \, U^{\dagger}_0 = \sum ^{\alpha \beta , s}_{kg}V^{\alpha \beta}_{kg}(t) \, f^{\alpha \dagger}_k f^{\beta}_g \left( a^s_q + a^{s\dagger}_{-q} \right)
	\end{split}
\end{equation}
where the interaction matrix elements $V^{\alpha \beta}_{kg}$ between the Floquet states have the form:
\begin{equation}
	\begin{split}
		& V^{\alpha \beta}_{kg}(t) = \sum_n \left[ \sum_m \langle \Psi ^{\alpha}_k(m+n)| \hat{V}^S | \Psi^{\beta}_g(m)\rangle \right] e^{\iota n\Omega t} \\
        & \equiv \sum_n V^{\alpha \beta}_{kg}(n) e^{\iota n\Omega t}
	\end{split}
    \end{equation}
And the integral over time for the interaction gives
\begin{equation}
	\begin{split}
		& \int _{-\infty}^t dt' \, \hat{V}_I(t')  = \sum ^{\rho, \sigma , s'}_{p,p'} f^{\rho \dagger}_p f^{\sigma }_{p'} \, V^{\rho \sigma}_{pp'}(n) \frac{a^{s'}_{q'} \, e^{\iota (\epsilon ^{\sigma \rho} _{p'p}+\omega ^{s'}_{q'}-n\Omega)t}}{\iota (\epsilon ^{\sigma \rho } _{p'p}+\omega ^{s'}_{q'}-n\Omega +\iota \eta)}\\
        & +\sum ^{\rho, \sigma , s'}_{p,p'} f^{\rho \dagger}_p f^{\sigma }_{p'} \, V^{\rho \sigma}_{pp'}(n)  \frac{a^{s' \dagger}_{-q'} \, e^{\iota (\epsilon ^{\sigma \rho }_{p'p}-\omega ^{s'}_{-q'}-n\Omega)t}}{\iota (\epsilon ^{\sigma \rho } _{p'p}-\omega ^{s'}_{-q'}-n\Omega+\iota \eta)} 
	\end{split}
\end{equation}
where $\eta$ is an arbitrarily small imaginary number introduced to regularize the integrals. Considering only the Cauchy principal value one obtains energy conserving delta functions as follows
\begin{equation}
	\begin{split}
		& U_0 \left( \int _{-\infty}^{t}dt' \, \hat{V}_I(t')  \right) U^{\dagger}_0\\
        & = \sum ^{\rho, \sigma , s'}_{p,p'} f^{\rho \dagger}_p f^{\sigma }_{p'}\, V^{\rho \sigma}_{pp'}(n) \;  a^{s'}_{q'} \; e^{\iota n\Omega t} \; \delta \left(\epsilon ^{\sigma \rho} _{p'p}+\omega ^{s'}_{q'}-n\Omega \right) \\
        & + \sum ^{\rho, \sigma , s'}_{p,p'} f^{\rho\dagger}_p f^{\sigma }_{p'} \, V^{\rho \sigma}_{pp'}(n) \; a^{s' \dagger}_{-q'} \; e^{\iota n\Omega t} \; \delta \left(\epsilon ^{\sigma \rho }_{p'p}-\omega ^{s'}_{-q'}-n\Omega \right) .
	\end{split}
\end{equation}
Going back to Eq. \ref{eq:dynpi}, the first commutator in the second term gives \small
\begin{equation}
	\begin{split}
		& \comm{\Theta^{ab}_k}{\hat{V}_I(t)} =  \sum ^{\alpha , \beta , s}_{g,g'}V^{\alpha \beta}_{gg'}(t) \,\left( a^s_q  + a^{s\dagger}_{-q}  \right) \comm{f^{a\dagger}_k f^b_k}{ f^{\alpha \dagger}_g f^{\beta}_{g'} } \\
        & = \sum_{k,g}^{\beta, s}   V^{b\beta }_{kg}(t) f^{a\dagger}_k f^{\beta}_g \left( a^s_q  + a^{s\dagger}_{-q} \right)  - \sum _{k,g}^{\alpha, s} V^{\alpha a}_{gk}(t) f^{\alpha \dagger}_g f^{b}_k \left( a^s_q  + a^{s\dagger}_{-q} \right)	.
     \end{split}
\end{equation}
\normalsize
In order to calculate the second commutator we need to trace out the bath operators from the product of $\comm{\Theta ^{ab}_k}{\hat{V}_I(t)} \int_{-\infty}^t dt' \, \hat{V}_I(t') $. We use Markovian approximation which assumes that the bath's response time ($\tau _p$) is much shorter than the dynamical time scale due to interaction Hamiltonian, $\tau _p \ll \frac{1}{|V|}$. Consequently, the phonon density matrix remains effectively constant and can be decoupled from the combined density operator as
 \begin{equation}\label{eq:markapp}
 	\hat{\rho} _I(t) \simeq \hat{\rho}^e_I(t) \otimes \hat{\sigma} ^p _I .
 \end{equation}
The decoupling implies $\langle aa^{\dagger}f^{\dagger}ff^{\dagger}f  \rangle \simeq \langle aa^{\dagger}  \rangle \langle f^{\dagger}ff^{\dagger}f \rangle$.\\
Now for the bath using zero temperature expectation values, i.e. $\langle a^s_q a^{s' \dagger}_{q'}\rangle = \delta ^{ss'}_{qq'}$ and assuming continuum energy spectrum with a constant density of states, $G$, for each momentum, i.e. $\sum ^s = G \int _0 ^{\infty} d\omega _q$, we obtain for the trace of  $\comm{\Theta ^{ab}_k}{\hat{V}_I(t)} \int_{-\infty}^t dt' \, \hat{V}_I(t') $,
\footnotesize
\begin{equation}\label{eq:bathtraceout}
	\begin{split}
		& Tr  \left \{ \comm{\Theta ^{ab}_k}{U_0 \hat{V}_I(t) U^{\dagger}_0} U_0\left( \int_{-\infty}^t dt' \, \hat{V}_I(t') \right) U^{\dagger}_0 \hat{\rho}_S \right \} _{ep}\\
       & = G \sum ^{\rho \sigma \beta }_{k,g,p,p'}    \, V^{\rho \sigma}_{pp'}(t) \;  V^{b\beta }_{kg}(t) \; Tr \left \{ f^{a\dagger}_k f^{\beta}_g  f^{\rho \dagger}_p f^{\sigma}_{p'} \, \hat{\rho}_S \right \}_e  \; \Theta \left(\epsilon ^{\sigma \rho}_{p'p} - \omega ^{s}_{q} -n\Omega \right) \\
       & - G \sum ^{\rho \sigma \beta }_{k,g,p,p'} V^{\rho \sigma}_{pp'}(t) \; V^{\alpha a}_{gk}(t) \; \langle f^{\alpha \dagger}_g f^{b}_k  f^{\rho \dagger}_p f^{\sigma}_{p'} \rangle \; \Theta \left(\epsilon ^{\sigma \rho}_{p'p} - \omega ^{s}_{q} -n\Omega \right)
	\end{split}
\end{equation}
\normalsize
Similar expression can be obtained for the term, $Tr \left \{ U_0 \left( \int _{-\infty}^t dt' \, \hat{V}_I(t') \right) \comm{\Theta ^{ab}_k}{U_0 \hat{V}_I(t) U^{\dagger}_0} \right \} _p $.\\
 While tracing out the electrons we assume the electronic system is non interacting and it has translational invariance so that
 \begin{equation}\label{eq:traceout}
 	\begin{split}
 		& \text{(i)} \;\;   \langle f^{a \dagger}_k f^{\beta}_g f^{\gamma \dagger}_p f^{\delta} _{p'} \rangle = \delta ^{\beta \gamma}_{gp} \langle f^{a \dagger}_k f^{\delta } _{p'} \rangle   \\
        &- \langle 
f^{a \dagger}_k f^{\delta} _{p'} \rangle \langle f^{\gamma \dagger}_p f^{\beta}_{g} \rangle + \langle f^{a \dagger}_k f^{\beta}_g \rangle \langle f^{\gamma \dagger}_p f^{\delta}_{p'} \rangle \\
			& \text{(ii)} \;\; \langle f^{\alpha \dagger}_k f^{\beta}_g \rangle = \delta _{kg} \langle f^{\alpha \dagger}_k f^{\beta}_k \rangle
 	\end{split}
\end{equation}
Using these expressions in Eq. \ref{eq:bathtraceout} and putting back to Eq. \ref{eq:dynpi} we obtain the rate equation as in Eq. \ref{eq:rateq}.
\section{Linearization procedure of master equation}\label{app:stepslin}
As argued in the main text, away from the degeneracy region only the valence bands are completely occupied which implies
\begin{equation}
	\begin{split}
		& \text{For $g$ such that $\Omega > 2\Delta _g$} \\
        & |\psi ^u_g \rangle = |v_g(-1)\rangle ,\;\;\;\;\;  |\psi ^d_g \rangle = |c_g(0)\rangle ; \\
        & \Pi ^{uu} _g =1, \;\;\; \Pi ^{dd}_g = 0 ,\;\;\; \Pi ^{ud}_g = 0 ;\\
        & \text{For $g$ such that $\Omega < 2\Delta _g$:} \\
        & |\psi ^u_g \rangle = |c_g (0)\rangle , \;\;\;\;\; |\psi ^d_g \rangle = |v_g(-1)\rangle ; \\
        & \Pi ^{uu} _g =0, \;\;\; \Pi ^{dd}_g = 1, \;\;\; \Pi ^{ud}_g = 0 .\\
	\end{split}
\end{equation}
where the states are written in frequency representation.\\
Putting $\Pi ^{\alpha \beta} _g$ values as input, Eq. \ref{eq:rateq} becomes linear in $\Pi ^{\alpha \beta}_k$ since we can ignore scatterings in between two degeneracy regions as its contribution is order of magnitude smaller.\\
Furthermore, using the state expressions we obtain for the interaction matrix elements,
\begin{equation}
	\begin{split}
		& \text{For $g$, such that $\Omega > 2\Delta _g $:}\\
        & V^{b u}_{kg}(n) V^{u \sigma}_{gk}(-n) = \langle \Psi ^{b}_k (n-1)| \hat{V}^S | v_g \rangle \langle v_g | \hat{V}^S | \Psi ^{\sigma}_k (n-1) \rangle \\
        & = \langle \Psi ^{b}_k (n-1) | \hat{S}^v | \Psi ^{\sigma}_k (n-1) \rangle .
	\end{split}
\end{equation}
Similarly all other matrix element combinations ($V^{\alpha \beta}_{kg}(n) V^{\gamma \delta }_{gk}(-n)$) for different $g$ points ($\Omega > 2\Delta _g \;\; \Omega < 2\Delta _g$) can be obtained.\\
The additional $\Theta$ function puts constraint on the values of $n$ that goes into the summation formula in the steady state rate equation. Consequently the contributions from $g: \Omega < 2\Delta _g$ become $O[A_0^2]$ and thus can be dropped. Incorporating all these, one obtains the $M_k$ and $\Lambda _k$ matrix elements as in Eq. \ref{eq:linrateq}.

\section{Steady state solutions and Current matrix elements}\label{app:curr}

 Using the expressions for Floquet states (Eq. \ref{eq:floqstates}) one can easily calculate the $M_k$ and $\Lambda _k$ matrix elements in Eq. \ref{eq:melements} to find the solution for Eq. \ref{eq:linrateq} in the form, $\Pi _k \equiv t_k \hat{I} + \vec{R}_k \cdot \vec{\sigma}$, given by
 \begin{equation}\label{eq:blocsoln}
  	\begin{split}
    	& t_k= \frac{1 - \frac{2(p-q)}{p}\frac{A_0^2|J^{\perp}_{0,k}|^2}{(p+q)^2+(2\epsilon _k)^2}}{1 + \frac{(p-q)^2}{pq}\frac{A_0^2|J^{\perp}_{0,k}|^2}{(p+q)^2+(2\epsilon _k)^2}}\\
        & \\
  		& R^z_k =\frac{2q+t_k(p-q)}{p+q} \frac{\frac{\Omega}{2}-\Delta _k}{2\epsilon _k} \\
        & \\
        & R^x_k = -\frac{A_0}{2\epsilon _k} \frac{(p+q)(2q+t_k(p-q))}{(p+q)^2+(2\epsilon _k)^2} \left( J^x_{0,k} - \frac{2 \epsilon _k}{p+q} J^y_{0,k}\right)\\
        & \\
        & R^y_k = -\frac{A_0}{2\epsilon _k} \frac{(p+q)(2q+t_k(p-q))}{(p+q)^2+(2\epsilon _k)^2} \left( J^y_{0,k} + \frac{2 \epsilon _k}{p+q} J^x_{0,k}\right) .
  	\end{split}
  \end{equation}
The current operator, up to first order in $A_0$, is given by
\begin{equation}\label{eq:currop}
 	\begin{split}
 		& \hat{J} =\sum ^m_k  \left( \begin{array}{cc}
 		c^{\dagger}_{k,m} & v^{\dagger}_{k,m}
 		\end{array} \right) \left( \vec{J}_{0,k} \cdot \vec{\sigma} \right) \left( \begin{array}{c}
 		c_{k,m} \\
        v_{k,m}
 		\end{array}\right) \\
        & +A_0 \sum ^m _k  \left( \begin{array}{cc}
 		c^{\dagger}_{k,m} & v^{\dagger}_{k,m}
 		\end{array} \right) \left( \vec{J}_{1,k} \cdot \vec{\sigma} \right) \left( \begin{array}{c}
 		c_{k,m\pm 1} \\
        v_{k,m\pm 1}
 		\end{array}\right) .
 	\end{split}
 \end{equation}
Using the Floquet state expressions (Eq. \ref{eq:floqstates}) along with the current operator definition in frequency space (Eq. \ref{eq:currop}), one can obtain the zero frequency component for the matrix elements of current operator, $\tilde{C}^{\alpha \beta}_k(0)$, in $\{ \hat{I}, \vec{\sigma}\}$ basis to be
\footnotesize
\begin{equation}\label{eq:currmatel}
	\begin{split}
    	& \tilde{C}^x_k(0) \equiv \frac{1}{2} \left(\tilde{C}^{ud}_k(0) + \tilde{C}^{du}_k(0) \right)\\
        & \simeq  \frac{ |x_k|}{1+|x_k|^2} \frac{J^x_{0,k} }{|J^{\perp}_{0,k}|} 2 J^z_{0,k}  - \frac{A_0}{1+|x_k|^2} \left( J^x_{1,k} \, sr_k  -J^y_{1,k} \, si_k\right) \\
        & \\
        & \tilde{C}^y_k (0)\equiv -\frac{1}{2\iota} \left(\tilde{C}^{up}_k(0) - \tilde{C}^{du}_k(0) \right)\\
        & \simeq \frac{|x_k|}{1+|x_k|^2}  \frac{J^y_{0,k} }{|J^{\perp}_{0,k}|} 2 J^z_{0,k}  - \frac{A_0}{1+|x_k|^2} \left( J^y_{1,k} \, sr_k  -J^x_{1,k} \, si_k\right)  \\
        & \\
        & \tilde{C}^z_k(0) \equiv \frac{1}{2}\left(\tilde{C}^{uu}_k(0) -\tilde{C}^{dd}_k(0) \right) \\
        & \simeq \frac{1}{1+|x_k|^2} \left[  J^z_{0,k} (1-|x_k|^2) + |x_k| \vec{J}_{0,k} \cdot \vec{J}_{1,k} \right]\\
        & \text{with} \;\; sr_k \equiv 1+ \frac{(J^x_{0,k})^2-(J^y_{0,k})^2}{|J^{\perp}_{0,k}|^2}|x_k|^2 \;\; \text{and} \;\;  si_k \equiv \frac{2J^x_{0,k} \, J^y_{0,k}}{|J^{\perp}_{0,k}|^2}|x_k|^2 .
	\end{split}
\end{equation}
\normalsize

The $\pm \Omega$ frequency components of the matrix elements of current operator, $\tilde{C}^{\alpha \beta}_k(\pm \Omega)$, are given by
\begin{equation}\label{eq:currmatelom}
	\begin{split}
		& \tilde{C}^{uu}_k (\Omega) = \frac{A_0}{2 \epsilon _k} \left( |J^{\perp}_{0,k}|^2 + J^z_{1,k} (\Omega - 2\Delta _k) \right) = \tilde{C}^{uu}_k(-\Omega) \\
        & \tilde{C}^{dd}_k(\pm \Omega) = -\tilde{C}^{uu}_k(\pm \Omega)\\
        & \tilde{C}^{ud}_k(-\Omega) = -\frac{J^x_{0,k} - \iota J^y_{0,k}}{1+|x_k|^2} \;\;\; \tilde{C}^{ud}_k(\Omega) = \frac{|x_k|^2}{1+|x_k|^2}(J^x_{0,k} - \iota J^y_{0,k}) \\
        & \tilde{C}^{du}_k(\Omega) =\left[ \tilde{C}^{ud}_k(-\Omega) \right]^* \;\;\;  \tilde{C}^{du}_k(-\Omega) =\left[ \tilde{C}^{ud}_k(\Omega) \right]^* .
	\end{split}
\end{equation}

\end{appendices}

\end{document}